# Complexity of Counting CSP with Complex Weights

Jin-Yi Cai
University of Wisconsin – Madison
jyc@cs.wisc.edu

Xi Chen
Columbia University
xichen@cs.columbia.edu

**Abstract**

We give a complexity dichotomy theorem for the counting Constraint Satisfaction Problem (#CSP in short) with complex weights. To this end, we give three conditions for its tractability. Let $\mathcal{F}$ be any finite set of complex-valued functions, then we prove that #CSP($\mathcal{F}$) is solvable in polynomial time if all three conditions are satisfied; and is #P-hard otherwise.

Our complexity dichotomy generalizes a long series of important results on counting problems: (a) the problem of counting graph homomorphisms is the special case when there is a single symmetric binary function in $\mathcal{F}$ [12, 5, 16, 7]; (b) the problem of counting directed graph homomorphisms is the special case when there is a single not-necessarily-symmetric binary function in $\mathcal{F}$ [10, 6]; and (c) the standard form of #CSP is when all functions in $\mathcal{F}$ take values in $\{0, 1\}$ [1, 13, 14].

# 1  Introduction

It is well known that if NP $\neq$ P, there is an infinite hierarchy of complexity classes between them [18]. However, for some broad classes of problems, a *complexity dichotomy* exists: every problem in the class is either in polynomial time or NP-hard. Such results include Schaefer's theorem [20], the dichotomy of Hell and Nešetřil for $H$-coloring [17], and some subclasses of the general constraint satisfaction problem (CSP in short) [9]. These developments lead to the following questions: How far can we push the envelope and show dichotomies for even broader classes of problems? Given a class of problems, what is the criterion that distinguishes the tractable problems from the intractable ones? The famous dichotomy conjecture by Feder and Vardi on decision CSP [15], which motivated much of the subsequent work, remains open to date. Now replacing NP with #P [23], both questions above can be asked for *counting* problems and in particular, the counting constraint satisfaction problem (#CSP in short).

In this paper, we study the complexity of #CSP in its most general form with complex weights. Let $D = \{1, \ldots, d\}$ be a finite set, called a *domain*. A weighted constraint language $\mathcal{F}$ over the domain $D$ is a finite set of complex-valued functions $\{f_1, \ldots, f_h\}$, where $f_i : D^{r_i} \to \mathbb{C}$ for some $r_i \geq 1$. The language $\mathcal{F}$ then defines the following counting constraint satisfaction problem, denoted by #CSP($\mathcal{F}$). The input consists of a tuple $\mathbf{x} = (x_1, \ldots, x_n)$ of variables over $D$ and a collection $I$ of tuples $(f, i_1, \ldots, i_r)$ in which $f$ is an $r$-ary function from $\mathcal{F}$ and $i_1, \ldots, i_r \in [n]$. It defines the following function $F_I$ over $\mathbf{x} \in D^n$:

$$F_I(\mathbf{x}) = \prod_{(f, i_1, \ldots, i_r) \in I} f(x_{i_1}, \ldots, x_{i_r}).$$

And the output of the problem is the following sum: $Z(I) = \sum_{\mathbf{x} \in D^n} F_I(\mathbf{x})$.

Various subclasses of #CSP have been studied intensively recently:

> **Counting graph homomorphisms**: This is the special case when the language $\mathcal{F}$ has a single symmetric binary function. A series of dichotomies have been discovered for functions with $\{0, 1\}$ weights by Dyer and Greenhill [12], nonnegative weights by Bulatov and Grohe [5], real weights by Goldberg, Grohe, Jerrum and Thurley [16], and complex weights by Cai, Chen and Lu [7].
>
> **Counting directed graph homomorphisms**: This is the special case when $\mathcal{F}$ has a single not-necessarily-symmetric binary function. In [11], Dyer, Goldberg and Paterson show a dichotomy for $\{0, 1\}$ functions that induce an acyclic graph when viewed as the adjacency matrix of a directed graph. Then Cai and Chen [6] give a dichotomy for all nonnegative binary functions.
>
> **The standard form of #CSP**: This is the special case when every function in $\mathcal{F}$ takes values in $\{0, 1\}$. Bulatov makes a breakthrough and proves a complexity dichotomy for this class (which we will refer to as unweighted #CSP). Later Dyer and Richerby give a simplified proof of his theorem and also prove the decidability of the dichotomy criterion in [13, 14]. It is then extended to include nonnegative and rational weights by Bulatov, Dyer, Goldberg, Jalsenius, Jerrum and Richerby [4], and nonnegative weights by Cai, Chen and Lu [8].

In this paper, we generalize all these results and prove a dichotomy for #CSP with complex weights:

**Theorem 1** (main). *Given any constraint language $\mathcal{F}$ with algebraic complex weights, #CSP($\mathcal{F}$) is either in polynomial time or #P-hard.*

To prove the dichotomy, we introduce three conditions on the language $\mathcal{F}$: the Block Orthogonality condition, the Mal'tsev condition and the Type Partition condition. We show that $\#\mathsf{CSP}(\mathcal{F})$ is #P-hard if $\mathcal{F}$ violates any of these three conditions; and give a polynomial-time algorithm that solves $\#\mathsf{CSP}(\mathcal{F})$ when $\mathcal{F}$ satisfies all three conditions.

**Proof Sketch**

The proof starts with the following framework for solving $\#\mathsf{CSP}(\mathcal{F})$. Let $I$ be an instance of $\#\mathsf{CSP}(\mathcal{F})$ and $F$ be the $n$-ary function it defines. We let $F^{[t]}$, for each $t \in [n]$, denote the following $t$-ary function:

$$F^{[t]}(x_1, \ldots, x_t) = \sum_{x_{t+1}, \ldots, x_n \in D} F(x_1, \ldots, x_t, x_{t+1}, \ldots, x_n).$$

In the discussion below, it is more convenient to consider $F^{[t]}$ as a $d^{t-1} \times d$ matrix when $t \geq 2$: the rows are indexed by $\mathbf{x} = (x_1, \ldots, x_{t-1})$; the columns are indexed by $i \in D$; and the $(\mathbf{x}, i)^{\text{th}}$ entry is $F^{[t]}(\mathbf{x}, i)$. In particular, we use $F^{[t]}(\mathbf{x}, *)$ to denote the $d$-dimensional row vector indexed by $\mathbf{x} \in D^{t-1}$.

In an ideal world, $Z(I)$ can be computed efficiently with the following *oracle*: We can send any tuple $\mathbf{x} \in D^{t-1}$ to the oracle, and it returns a $d$-dimensional vector $\mathbf{v}$ that is linearly dependent with $F^{[t]}(\mathbf{x}, *)$. Here either $\mathbf{v} = \mathbf{0}$ if $F^{[t]}(\mathbf{x}, *) = \mathbf{0}$; or $\mathbf{v}$ has its first non-zero entry normalized to 1 so that it is unique.

With the help of such a powerful oracle, we can compute $Z(I)$ as follows. From $Z(I) = \sum_{a \in D} F^{[1]}(a)$ it suffices to compute $F^{[1]}(a)$ for each $a \in D$. Pick any $a_1 \in D$, and we send it to the oracle. The oracle returns a vector $\mathbf{v}$ that is linearly dependent with $F^{[2]}(a_1, *)$. If $\mathbf{v} = \mathbf{0}$ then $F^{[1]}(a_1) = \sum_{b \in D} F^{[2]}(a_1, b) = 0$. Otherwise, let $a_2 \in D$ be the index of the first non-zero entry of $\mathbf{v}$, with $v_{a_2} = 1$. Then we have

$$F^{[1]}(a_1) = \sum_{b \in D} F^{[2]}(a_1, b) = F^{[2]}(a_1, a_2) \cdot \sum_{b \in D} v_b,$$

where the last equation follows from the assumption that $F^{[2]}(a_1, *)$ and $\mathbf{v}$ are linearly dependent. This reduces the computation of $F^{[1]}(a_1)$ to $F^{[1]}(a_1, a_2)$. Next we send $(a_1, a_2)$ to the oracle. Either the vector $\mathbf{w}$ we get back from the oracle is $\mathbf{0}$, in which case $F^{[2]}(a_1, a_2) = 0$; or we can use $\mathbf{w}$ to further reduce the computation of $F^{[1]}(a_1)$ to $F^{[3]}(a_1, a_2, a_3)$ for some appropriate $a_3 \in D$. Repeating this process for $n-1$ rounds, then it suffices to compute $F^{[n]}(a_1, a_2, \ldots, a_n)$ for some appropriate $a_2, \ldots, a_n \in D$. This gives an efficient algorithm for computing $F^{[1]}(a_1)$ because $F = F^{[n]}$ can be evaluated efficiently using $I$.

As a result, we can solve $\#\mathsf{CSP}(\mathcal{F})$ efficiently with such an oracle. It turns out that almost the whole proof of Theorem 1 is trying to understand *how* and *when* we can efficiently implement this oracle. Note that we essentially need to "collect" the following huge amount of information: For each $t \in [n]$, we need to compute a set of pairwise linearly independent (and normalized) $d$-dimensional vectors $\mathbf{v}^{[t,1]}, \ldots, \mathbf{v}^{[t,s_t]}$ for some $s_t \geq 0$, so that every nonzero row vector $F^{[t]}(\mathbf{x}, *)$ is linearly dependent with one of them. Also for each vector $\mathbf{v}^{[t,j]}$, we need to know the set of $\mathbf{x} \in D^{t-1}$, denoted by $S^{[t,j]} \subseteq D^{t-1}$, such that $F^{[t]}(\mathbf{x}, *)$ is non-zero and linearly dependent with $\mathbf{v}^{[t,j]}$. Two difficulties arise. First, note that in general an $m \times d$ matrix may have as many as $m$ pairwise linearly independent row vectors. Thus in general, we may need to keep track of exponentially many $\mathbf{v}^{[t,j]}$'s. Second, for each $\mathbf{v}^{[t,j]}$, $S^{[t,j]}$ is in general exponential in $t$.

To overcome the first difficulty, we drew inspiration from the recent dichotomy theorems for counting graph homomorphisms with real [16] and complex weights [7]. In both dichotomies those tractable cases are closely related to matrices in which every two row vectors are either linearly dependent or orthogonal e.g., the Hadamard matrices and the so-called discrete unitary matrices [7]. This inspires us to introduce the first necessary condition for tractability: the Block Orthogonality condition. It requires that for any $F$ defined by an instance of #CSP($\mathcal{F}$) and for any $t \in [n]$, every two row vectors of $F^{[t]}$ are either linearly dependent or orthogonal; Otherwise #CSP($\mathcal{F}$) is #P-hard. Actually a more stringent requirement than orthogonality must hold (as the word "block" suggests); Otherwise #CSP($\mathcal{F}$) is #P-hard. See the formal definition in Section 3.1. Assume $\mathcal{F}$ satisfies the Block Orthogonality condition. Then we know for sure that each $F^{[t]}$ has at most $d$ pairwise linearly independent (and indeed pairwise orthogonal) row vectors.

To overcome the second difficulty, we need some of the powerful techniques developed for the unweighted #CSP [1, 13]. One of the tools used there is the notion of Mal'tsev polymorphism from universal algebra (see Section 2.8). In [13] Dyer and Richerby introduce a succinct representation, called a *witness function*, for any set $\Phi \subseteq D^n$ that has a Mal'tsev polymorphism $\varphi$. A witness function of a set $\Phi \subseteq D^n$ is of linear size in $n$, the arity of $\Phi$, and essentially contains all the information about $\Phi$. In particular, with a witness function one can decide whether a given tuple $\mathbf{x} \in D^n$ belongs to $\Phi$ efficiently. From here it is only natural to ask whether the sets $S^{[t,j]}$ associated with each $\mathbf{v}^{[t,j]}$ have a Mal'tsev polymorphism. This is where we introduce the second necessary condition: the Mal'tsev condition. Roughly speaking, it requires all the sets $S^{[t,j]} \subseteq D^{t-1}$, defined from all $F, t$ and $j$, to share a common Mal'tsev polymorphism $\varphi$; Otherwise the problem #CSP($\mathcal{F}$) is #P-hard.

We can now refine the plan of implementing the oracle as follows. Assume the language $\mathcal{F}$ satisfies both the Block Orthogonality condition and the Mal'tsev condition. Then given any input instance $I$ of #CSP($\mathcal{F}$) which defines an $n$-ary function $F$, we compute for each $t : 2 \le t \le n$,

(a) A set of (at most $d$) pairwise orthogonal and normalized $d$-dimensional vectors $\mathbf{v}^{[t,1]}, \ldots, \mathbf{v}^{[t,s_t]}$, for some $s_t \ge 0$, such that every nonzero $F^{[t]}(\mathbf{x}, *)$ is linearly dependent with one of them.

(b) A witness function $\omega^{[t,j]}$ for each set $S^{[t,j]}$, which can be used to decide membership efficiently.

So the only algorithmic problem left is, how (and when) can we compute these objects efficiently?

To this end, we start with $t = n$ and $F = F^{[n]}$. First, by using the Mal'tsev condition and a beautiful algorithm of Dyer and Richerby [13], we can construct efficiently a witness function $\omega$ for $R \subseteq D^n$ where $\mathbf{x} \in R$ if and only if $F(\mathbf{x}) \ne 0$. With $\omega$, it is also easy to construct a witness function $\omega'$ for $R' \subseteq D^{n-1}$, the projection of $R$ on its first $n-1$ coordinates. We are getting closer because by the definition of $S^{[n,j]}$, $R'$ is exactly the union of the $s_n$ pairwise disjoint sets $S^{[n,1]}, \ldots, S^{[n,s_n]} \subseteq D^{n-1}$. Skipping some technical details, what we need boils down to the following *splitting* operation over witness functions:

> Let $\Phi \subseteq D^n$ be a nonempty set, and let $\Psi_1, \ldots, \Psi_s$ be an $s$-way partition of $\Phi$, for some $s \in [d]$: The $\Psi_i$'s are nonempty, pairwise disjoint, and satisfy $\Phi = \Psi_1 \cup \cdots \cup \Psi_s$. Assume that $\varphi$ is a Mal'tsev polymorphism of $\Phi$ and all the $\Psi_i$'s. At the beginning, we have completely no information about the $\Psi_i$'s, not even the number $s$ of the $\Psi_i$'s, though we do know that $s \in [d]$. The only resources we have are a witness function $\omega$ for $\Phi$ and a black box to query: We can send any $\mathbf{x} \in \Phi$ to the black box and it returns the unique index $k \in [s]$ such that $\mathbf{x} \in \Psi_k$. The question is: Can we use $\omega$ and the black box to compute $s \in [d]$ as well as a witness function $\omega_k$ for each $\Psi_k$ in polynomial time and only using polynomially many queries?

In general, we do not know how to implement the splitting operation above efficiently. However, if $\Phi$ and $\Psi_1, \ldots, \Psi_s$ satisfy the so-called partition condition (see the definition in Section 3.1) then we present an algorithm that computes $s \in [d]$ as well as a witness function $\omega_k$ for each $\Psi_k$, in polynomial time and using polynomially many queries. This finally brings us to the last condition: the Type Partition condition. It turns out that this condition is also necessary for tractability. If $\mathcal{F}$ violates it, then $\#\mathsf{CSP}(\mathcal{F})$ is #P-hard. Roughly speaking, the Type Partition condition requires that whenever we need to apply the splitting operation, $\Phi$ and $\Psi_1, \ldots, \Psi_s$ satisfy the partition condition so that our algorithm works. In particular, it requires $R'$ and $S^{[n,1]}, \ldots, S^{[n,s_n]}$ to satisfy the partition condition and thus, we can apply the splitting operation to construct a witness function for each $S^{[n,j]}$ using $\omega'$. The proof showing that the Type Partition condition is necessary (Section 5) and the algorithm for the splitting operation assuming the partition condition (Section 7.3) are among the most challenging in the paper. Moreover, with the help of the splitting operation and the Type Partition condition, we can inductively construct a witness function for each $S^{[t,j]}$ from $t = n$ to $2$. Therefore, we get an efficient implementation of the oracle and thus, the problem $\#\mathsf{CSP}(\mathcal{F})$ is in polynomial time when all three necessary conditions are satisfied. This finishes the proof of Theorem 1.

## 2 Preliminaries

### 2.1 Notation

For convenience, we let $\mathbb{C}$ denote the set of algebraic complex numbers throughout the paper and usually refer to them simply as complex numbers when it is clear from the context.

Let $D = \{1, 2, \ldots, d\}$ be a finite set, called a *domain*. Let $F: D^n \to \mathbb{C}$ be an $n$-ary complex function. We use $\mathsf{Im}(F)$ to denote the image of $F$, i.e.,

$$\mathsf{Im}(F) = \{c \in \mathbb{C} : c = F(\mathbf{x}) \text{ for some } \mathbf{x} \in D^n\}.$$

Given a finite set $\mathcal{F} = \{F_1, \ldots, F_h\}$ of functions, we use $\mathsf{Im}(\mathcal{F})$ to denote the image of $\mathcal{F}$:

$$\mathsf{Im}(\mathcal{F}) = \mathsf{Im}(F_1) \cup \cdots \cup \mathsf{Im}(F_h).$$

Given $F: D^n \to \mathbb{C}$, we use $F^{[t]}$, for each $t \in [n]$, to denote the following $t$-ary function:

$$F^{[t]}(x_1, \ldots, x_t) = \sum_{x_{t+1}, \ldots, x_n \in D} F(x_1, \ldots, x_t, x_{t+1}, \ldots, x_n).$$

Note that $F^{[n]} = F$. We also use $|F|$ to denote the non-negative function defined as follows:

$$|F| : \mathbf{x} \mapsto |F(\mathbf{x})|, \quad \text{for all } \mathbf{x} \in D^n.$$

Given $\mathcal{F} = \{F_1, \ldots, F_h\}$, we use $|\mathcal{F}|$ to denote $\{|F_1|, \ldots, |F_h|\}$.

Given $F: D^n \to \mathbb{C}$ where $n \geq 2$, in certain situations we consider $F$ as a matrix with exponentially many rows but only $d$ columns. We let $\mathbf{M}_F$ denote the following $d^{n-1} \times d$ matrix: The rows of $\mathbf{M}_F$ are indexed by $\mathbf{x} = (x_1, \ldots, x_{n-1}) \in D^{n-1}$; the columns are indexed by $x_n \in D$; and the $(\mathbf{x}, x_n)^{\text{th}}$ entry

$$M_F(\mathbf{x}, x_n) = F(\mathbf{x}, x_n) = F(x_1, \ldots, x_{n-1}, x_n).$$

For any $\mathbf{x} \in D^{n-1}$, we use $F(\mathbf{x}, *)$ to denote the $d$-dimensional vector whose $i^{\text{th}}$ entry is $F(\mathbf{x}, i)$. We use $|F(\mathbf{x}, *)|$ to denote the $d$-dimensional non-negative vector whose $i^{\text{th}}$ entry is $|F(\mathbf{x}, i)|$.

Given an $m \times n$ matrix $\mathbf{M}$, we use $\mathbf{M}(i, *)$ to denote the $i^{\text{th}}$ row vector of $\mathbf{M}$.

Given two vectors $\mathbf{x}$ and $\mathbf{y} \in \mathbb{C}^d$, we say they are orthogonal if

$$\sum_{i \in [d]} x_i \cdot \overline{y_i} = 0.$$

Given $\mathbf{x} \in D^n$ and $\ell \in [n]$, we use $\mathsf{Pr}_{[\ell]} \mathbf{x}$ to denote its prefix of length $\ell$.

Let $\Phi \subseteq D^n$ be an $n$-ary relation. For each $\ell \in [n]$, we let $\mathsf{Pr}_\ell \Phi \subseteq D$ denote the *projection* of $\Phi$ on the $\ell^{\text{th}}$ coordinate: $a \in \mathsf{Pr}_\ell \Phi$ if and only if there is an $\mathbf{x} \in \Phi$ such that $x_\ell = a$. We call $\mathbf{x}$ a *witness* for $a$ at the $\ell^{\text{th}}$ coordinate, or simply a witness for the pair $(\ell, a)$. We also use $\mathsf{Pr}_{[\ell]} \Phi \subseteq D^\ell$ to denote the projection of $\Phi$ on the first $\ell$ coordinates: $\mathbf{y} \in \mathsf{Pr}_{[\ell]} \Phi$ if and only if there exists an $\mathbf{x} \in \Phi$ such that $\mathbf{y} = \mathsf{Pr}_{[\ell]} \mathbf{x}$.

Given a vector $\mathbf{a} \in D^\ell$ for some $\ell \in [n]$, we let $\Phi(\mathbf{a}, *) = \Phi(a_1, \ldots, a_\ell, *)$ denote the relation on $n - \ell$ variables with the first $\ell$ variables fixed to $\mathbf{a}$: $\mathbf{y} \in \Phi(\mathbf{a}, *)$ if and only if $\mathbf{a} \circ \mathbf{y} \in \Phi$.

Given a permutation $\pi$ over $[n]$, we let $\Phi_\pi$ denote the $n$-ary relation such that $\mathbf{x} \in \Phi_\pi$ if and only if

$$(x_{\pi(1)}, \ldots, x_{\pi(n)}) \in \Phi.$$

Finally we use $\leq_T$ to denote polynomial-time Turing reductions between problems, and $\equiv_T$ to denote equivalence under polynomial-time Turing reductions.

## 2.2 Counting CSP with Algebraic Weights

Let $D = [d]$ be a domain, and $\mathcal{F} = \{F_1, \ldots, F_h\}$ be a finite set of complex functions over $D$. They define the following problem denoted by $\#\mathsf{CSP}(\mathcal{F})$. An input instance of the problem consists of a finite set of variables $x_1, \ldots, x_n$ over $D$ and a finite multiset $I$ of tuples $(F, i_1, \ldots, i_r)$ in which $F$ is an $r$-ary function in $\mathcal{F}$ and $i_1, \ldots, i_r \in [n]$. It defines the following $n$-ary function $F_I$ over $\mathbf{x} = (x_1, \ldots, x_n) \in D^n$:

$$F_I(\mathbf{x}) = \prod_{(F, i_1, \ldots, i_r) \in I} F(x_{i_1}, \ldots, x_{i_r}).$$

The output of the problem is then the following exponential sum:

$$Z(F_I) = \sum_{\mathbf{x} \in D^n} F_I(\mathbf{x}).$$

When $\mathcal{F} = \{F\}$ has only one function, we also use $\#\mathsf{CSP}(F)$ to denote $\#\mathsf{CSP}(\mathcal{F})$ for convenience.

To complete the definition of $\#\mathsf{CSP}(\mathcal{F})$, we need to specify the model of algebraic number computation, i.e., how the numbers in $\mathcal{F}$ and the output $Z(F_I)$ are encoded. We can take any reasonable model, e.g., the one used earlier in [22, 21, 19]. This issue of computation model does not seem central to this paper because when the complexity of $\#\mathsf{CSP}(\mathcal{F})$ is concerned, $\mathcal{F}$ is fixed and considered as a constant. The input size only depends on $n$, the number of variables, and $|I|$, the number of tuples in $I$.

Given $D$ and $\mathcal{F}$, we can also define the following problem, denoted $\mathsf{COUNT}(\mathcal{F})$: The input is a pair $(I, c)$, where $I$ is an input instance of $\#\mathsf{CSP}(\mathcal{F})$ and $c$ is an algebraic complex number. Let $x_1, \ldots, x_n$ be the variables over $D$ in $I$, then the output is the number of $\mathbf{x} = (x_1, \ldots, x_n) \in D^n$ such that

$$F_I(\mathbf{x}) = c, \quad \text{where } F_I \text{ is the } n\text{-ary function defined by } I.$$

It turns out that $\mathsf{COUNT}(\mathcal{F})$ and $\#\mathsf{CSP}(\mathcal{F})$ are equivalent under polynomial-time Turing reductions:

**Lemma 1.** $\mathsf{COUNT}(\mathcal{F}) \equiv_\mathsf{T} \#\mathsf{CSP}(\mathcal{F})$.

*Proof.* Let $\mathsf{Im}(\mathcal{F}) = \{c_1, \ldots, c_k\}$ where $k = |\mathsf{Im}(\mathcal{F})|$ is considered as a constant because $\mathcal{F}$ is fixed. Let $I$ be an input instance of $\#\mathsf{CSP}(\mathcal{F})$ over $n$ variables $\mathbf{x} \in D^n$ with $m = |I|$, and let $F$ be the $n$-ary function that $I$ defines. First of all, we can compute the following set of numbers in time polynomial in $m$:

$$C_m = \left\{ c_1^{\ell_1} \cdots c_k^{\ell_k} : \ell_1, \ldots, \ell_k \text{ are non-negative integers and } \ell_1 + \cdots + \ell_k = m \right\},$$

since $k$ is a constant. It then follows from the definition of $F$ that $F(\mathbf{x}) \in C_m$ for all $\mathbf{x} \in D^n$.

For each $c \in C_m$, we let $N_c$ denote the number of $\mathbf{x} \in D^n$ such that $F(\mathbf{x}) = c$, then we have

$$Z(F) = \sum_{\mathbf{x} \in D^n} F(\mathbf{x}) = \sum_{c \in C_m} c \cdot N_c.$$

This immediately gives us a polynomial-time reduction from $\#\mathsf{CSP}(\mathcal{F})$ to $\mathsf{COUNT}(\mathcal{F})$.

We prove the other direction: Given any $I$, we use a subroutine for $\#\mathsf{CSP}(\mathcal{F})$ to compute $N_c$ for all $c \in C_m$. For this purpose, we let $C'_m = C_m - \{0\}$ and let $s = |C'_m|$ which is polynomial in $m$. We build from $I$ the following instances $I_1, \ldots, I_s$: to get $I_\ell$, $\ell \in [s]$, we make $\ell$ copies of each tuple in $I$ and thus, $I_1 = I$ and $|I_\ell| = \ell \cdot |I|$. We also let $F_\ell$ denote the $n$-ary function defined by $I_\ell$.

By the construction of $I_\ell$, it is easy to see that $F_\ell(\mathbf{x}) = (F(\mathbf{x}))^\ell$ for all $\mathbf{x} \in D^n$ and thus,

$$Z(F_\ell) = \sum_{\mathbf{x} \in D^n} F_\ell(\mathbf{x}) = \sum_{c \in C_m} c^\ell \cdot N_c = \sum_{c \in C'_m} c^\ell \cdot N_c, \quad \text{for each } \ell = 1, \ldots, s.$$

The left hand side of the equations can be obtained by calling a subroutine for $\#\mathsf{CSP}(\mathcal{F})$ on $I_\ell$. We can then solve the Vandermonde system above to get $N_c$ for each $c \in C'_m$. If $0 \in C_m$, we can also derive $N_0$ using the fact that the sum of all the $N_c$'s, $c \in C_m$, is $d^n$. This finishes the proof of the lemma. $\square$

In certain situations the problem $\mathsf{COUNT}(\mathcal{F})$ is easier to use than $\#\mathsf{CSP}(\mathcal{F})$. For example, we use it to prove the following lemma.

**Lemma 2.** $\#\mathsf{CSP}(|\mathcal{F}|) \leq_\mathsf{T} \#\mathsf{CSP}(\mathcal{F})$.

*Proof.* By Lemma 1, it suffices to show $\mathsf{COUNT}(|\mathcal{F}|) \leq_\mathsf{T} \mathsf{COUNT}(\mathcal{F})$. We let $\mathsf{Im}(\mathcal{F}) = \{c_1, \ldots, c_k\}$ where $k$ is considered as a constant because the set $\mathcal{F}$ of functions is fixed.

Let $I$ be an input instance of $\#\mathsf{CSP}(|\mathcal{F}|)$, and $F$ be the $n$-ary non-negative function it defines. Let $a$ be a non-negative number, and we need to compute the number of $\mathbf{x} \in D^n$ such that $F(\mathbf{x}) = a$.

From $I$, it is natural to construct an input instance $I'$ of $\#\mathsf{CSP}(\mathcal{F})$ by simply replacing the function $|F_i|$ in each tuple of $I$ with its corresponding function $F_i$ in $\mathcal{F}$. Let $F'$ denote the function that $I'$ defines then it is clear that $F(\mathbf{x}) = |F'(\mathbf{x})|$ for all $\mathbf{x} \in D^n$.

Let $m = |I| = |I'|$, then we can compute

$$C_m = \left\{ c_1^{\ell_1} \cdots c_k^{\ell_k} : \ell_1, \ldots, \ell_k \text{ are non-negative integers and } \ell_1 + \cdots + \ell_k = m \right\}$$

in time polynomial in $m$ because $k$ is a constant.

From the definitions of $C_m$ and $F'$, we have $F'(\mathbf{x}) \in C_m$ for all $\mathbf{x} \in D^n$. As a result,

$$\left[ \text{the number of } \mathbf{x} \text{ such that } F(\mathbf{x}) = a \right] = \sum_{c \in C_m: |c| = a} \left[ \text{the number of } \mathbf{x} \text{ such that } F'(\mathbf{x}) = c \right],$$

and the right hand side can be computed efficiently, because the number of such $c$ can be no more than $|C_m|$ and the term for each $c$ can be evaluated by calling a subroutine for $\mathsf{COUNT}(\mathcal{F})$.

This finishes the proof of the lemma. $\square$

## 2.3 Row Representation

Let $\mathbf{M}$ be an $m \times n$ complex matrix. It induces the following equivalence relation $\sim_\mathbf{M}$ over

$$\left\{ \ell \in [m] : \mathbf{M}(\ell, *) \neq \mathbf{0} \right\},$$

i.e., the set of nonzero rows of $\mathbf{M}$:

$$\ell \sim_\mathbf{M} \ell' \iff \mathbf{M}(\ell, *) \text{ and } \mathbf{M}(\ell', *) \text{ are linearly dependent over } \mathbb{C}.$$

We say that $\mathcal{S} = \{(S_1, \mathbf{v}_1), \ldots, (S_k, \mathbf{v}_k)\}$, for some integer $k \geq 0$, is the *row representation* of $\mathbf{M}$ if

1. $S_1, \ldots, S_k \subseteq [m]$ are the equivalence classes of the equivalence relation $\sim_\mathbf{M}$; and

2. For each $i \in [k]$, $\mathbf{v}_i$ is a nonzero $n$-dimensional vector with its first nonzero entry being 1, and is linearly dependent with $\mathbf{M}(\ell, *)$, for all $\ell \in S_i$. (By the definition of $\sim_\mathbf{M}$, $\mathbf{v}_i$ exists and is unique.)

We will refer to $\mathbf{v}_i$ as the *representative row vector* for the equivalence class $S_i$.

From the definition, it is easy to see that $S_i$ is nonempty for all $i$; the $S_i$'s are pairwise disjoint;

$$S_1 \cup \cdots \cup S_k = \left\{ \ell \in [m] : \mathbf{M}(\ell, *) \neq \mathbf{0} \right\};$$

for all $i \neq j$, $\mathbf{v}_i$ and $\mathbf{v}_j$ are linearly independent. Clearly every matrix has a unique row representation.

In general, the row representation $\mathcal{S}$ of an $m \times n$ matrix $\mathbf{M}$ may consist of as many as $m$ pairs. But if it is known that every two rows of $\mathbf{M}$ are either linearly dependent or orthogonal, then the number of pairs in its row representation cannot exceed $n$.

For a non-negative $\mathbf{M}$, we say it is *block-rank*-1 if its row representation $\mathcal{S} = \{(S_1, \mathbf{v}_1), \ldots, (S_k, \mathbf{v}_k)\}$ has the property that for all $i \neq j \in [k]$, the two vectors $\mathbf{v}_i$ and $\mathbf{v}_j$ have distinct positive entries.

We next extend the notion of row representations to functions. Given any $F : D^n \to \mathbb{C}$ with $n \geq 2$, we define the following equivalence relation $\sim_F$ over $\{\mathbf{x} \in D^{n-1} : F(\mathbf{x}, *) \neq \mathbf{0}\}$:

$$\mathbf{x} \sim_F \mathbf{y} \iff F(\mathbf{x}, *) \text{ and } F(\mathbf{y}, *) \text{ are linearly dependent over } \mathbb{C}.$$

Similarly, we say

$$\mathcal{S} = \{(S_1, \mathbf{v}_1), \ldots, (S_k, \mathbf{v}_k)\}, \quad \text{where } S_i \subseteq D^{n-1} \text{ for all } i \in [k],$$

is the row representation of $F$ if $\mathcal{S}$ is the row representation of the $d^{n-1} \times d$ matrix $\mathbf{M}_F$. (Equivalently, we have $S_1, \ldots, S_k$ are the equivalence classes of $\sim_F$; and for each $i \in [k]$, $\mathbf{v}_i$ is a non-zero $d$-dimensional vector with its first non-zero entry being 1, and is linearly dependent with $F(\mathbf{x}, *)$, $\mathbf{x} \in S_i$.)

Finally, we also call $F$ a *block-rank-1 function*, if $\mathbf{M}_{|F|}$ is a block-rank-1 matrix. (Equivalently, for all $\mathbf{x}, \mathbf{y} \in D^{n-1}$ with $F(\mathbf{x}, *)$ and $F(\mathbf{y}, *)$ being nonzero, the two non-negative vectors $|F(\mathbf{x}, *)|$ and $|F(\mathbf{y}, *)|$ either are linearly dependent or share no common positive entry.)

## 2.4 The Block-Rank-1 Condition

We need the following dichotomy from Bulatov and Grohe [5]. Let $\mathbf{A}$ denote a symmetric $d \times d$ non-negative matrix with algebraic entries. It defines the following graph homomorphism problem, denoted by EVAL($\mathbf{A}$): The input is an undirected graph $G = (V, E)$ with $V = [n]$, and the output is

$$Z_{\mathbf{A}}(G) = \sum_{x_1, \ldots, x_n \in [d]} \left( \prod_{ij \in E} A(x_i, x_j) \right).$$

In the language of #CSP, EVAL($\mathbf{A}$) is the same as #CSP($F$) with $F(i, j) = A(i, j)$ for all $i, j \in [d]$.

**Theorem 2.** *Let $\mathbf{A}$ be a symmetric and non-negative square matrix with algebraic entries, then EVAL($\mathbf{A}$) is in polynomial time if $\mathbf{A}$ is block-rank-1; and is #P-hard otherwise.*

We can also extend the definitions of EVAL($\mathbf{A}$) and $Z_{\mathbf{A}}(\cdot)$ to any square matrix $\mathbf{A}$ over $\mathbb{C}$. The input of EVAL($\mathbf{A}$) is now a *directed* graph $G = (V, E)$, and the output is

$$Z_{\mathbf{A}}(G) = \sum_{x_1, \ldots, x_n \in [d]} \left( \prod_{\overrightarrow{ij} \in E} A(x_i, x_j) \right).$$

The following lemma will be useful later in the proof:

**Lemma 3.** *Let $\mathbf{A}$ be a square (though not necessarily symmetric) matrix with algebraic complex entries. If $|\mathbf{A}|$ is not block-rank-1, then EVAL($\mathbf{A}$) is #P-hard.*

*Proof.* By Lemma 2, it suffices to show that EVAL($|\mathbf{A}|$) is #P-hard. To this end we use $\mathbf{B}$ to denote the symmetric and non-negative $d \times d$ matrix $|\mathbf{A}||\mathbf{A}|^T$, where the $(i, j)^{\text{th}}$ entry of $\mathbf{B}$ is

$$B(i, j) = \sum_{k \in [d]} |A(i, k)| \cdot |A(j, k)|.$$

From the definition of $\mathbf{B}$, we claim EVAL($\mathbf{B}$) $\leq_T$ EVAL($|\mathbf{A}|$). This is because given any undirected graph $G = (V, E)$ of EVAL($\mathbf{B}$), we can construct a new directed graph $G' = (V', E')$ with

$$V' = \left\{ x_v, x_e : v \in V \text{ and } e \in E \right\} \quad \text{and} \quad E' = \left\{ \overrightarrow{x_u x_e}, \overrightarrow{x_v x_e} : e = uv \in E \right\}.$$

It is then easy to check that $Z_{\mathbf{B}}(G) = Z_{|\mathbf{A}|}(G')$ from which the reduction follows. On the other hand, by Cauchy-Schwarz, if $|\mathbf{A}|$ is not block-rank-1, neither is $\mathbf{B}$. Then it follows from Theorem 2 that $\mathsf{EVAL}(\mathbf{B})$ is #P-hard, and so is $\mathsf{EVAL}(|\mathbf{A}|)$. This finishes the proof. □

Next, we use Theorem 2 to give a useful #P-hardness lemma for #CSP with a single complex-valued function. The idea is similar to the proof of Lemma 3 above. Let $D = [d]$ be a domain, then

**Lemma 4** (The Block-Rank-1 Condition). *Let $F : D^r \to \mathbb{C}$ be any algebraic complex function with arity $r \geq 2$. If $F$ is not a block-rank-1 function, then $\#\mathsf{CSP}(F)$ is #P-hard.*

*Proof.* First of all, by Lemma 2 it suffices to show that $\#\mathsf{CSP}(|F|)$ is #P-hard.

To finish the proof, we construct a symmetric and non-negative matrix $\mathbf{A}$ from $|F|$ such that

$$\mathsf{EVAL}(\mathbf{A}) \leq_T \#\mathsf{CSP}(|F|) \tag{1}$$

and then use Theorem 2 to show that $\mathsf{EVAL}(\mathbf{A})$ is #P-hard.

To this end, we define the following matrix $\mathbf{A}$: its rows and columns are indexed by $\mathbf{x} \in D^{r-1}$, and

$$A(\mathbf{x}, \mathbf{y}) = \sum_{i \in D} |F(\mathbf{x}, i)| \cdot |F(\mathbf{y}, i)|.$$

It is clear that $\mathbf{A}$ is both symmetric and non-negative.

Moreover, given any undirected graph $G = (V, E)$ of $\mathsf{EVAL}(\mathbf{A})$, we construct the following instance $I$ of $\#\mathsf{CSP}(|F|)$: it has the following $(r-1)|V| + |E|$ variables

$$x_{v,1}, \ldots, x_{v,r-1}, y_e, \quad \text{for each } v \in V \text{ and } e \in E.$$

For every $e = uv \in E$, we add the following two tuples to $I$:

$$\Big(|F|, x_{u,1}, \ldots, x_{u,r-1}, y_e\Big) \quad \text{and} \quad \Big(|F|, x_{v,1}, \ldots, x_{v,r-1}, y_e\Big).$$

From the construction of $I$ and the definition of $\mathbf{A}$ from $|F|$, it is easy to check that

$$Z_{\mathbf{A}}(G) = Z(F_I), \quad \text{where } F_I \text{ is the function that } I \text{ defines.}$$

This gives us a polynomial-time reduction from $\mathsf{EVAL}(\mathbf{A})$ to $\#\mathsf{CSP}(|F|)$.

Finally, we show that if $F$ is not block-rank-1, then $\mathbf{A}$ is not a block-rank-1 matrix, and by Theorem 2, $\mathsf{EVAL}(\mathbf{A})$ is #P-hard. Because $F$ is not block-rank-1, we know there are two vectors $\mathbf{x}, \mathbf{y} \in D^{r-1}$ such that $|F(\mathbf{x}, *)|$ and $|F(\mathbf{y}, *)|$ share at least one common positive entry but are linearly independent. This implies that all the following four entries of $\mathbf{A}$ are positive:

$$A(\mathbf{x}, \mathbf{x}), \ A(\mathbf{x}, \mathbf{y}) = A(\mathbf{y}, \mathbf{x}), \ A(\mathbf{y}, \mathbf{x}) > 0,$$

but by Cauchy-Schwarz, we have

$$A(\mathbf{x}, \mathbf{y}) \cdot A(\mathbf{y}, \mathbf{x}) = \left(\sum_{i \in D} |F(\mathbf{x}, i)| \cdot |F(\mathbf{y}, i)|\right)^2 < \left(\sum_{i \in D} |F(\mathbf{x}, i)|^2\right)\left(\sum_{i \in D} |F(\mathbf{y}, i)|^2\right) = A(\mathbf{x}, \mathbf{x}) \cdot A(\mathbf{y}, \mathbf{y}).$$

Therefore, $\mathbf{A}$ is not a block-rank-1 matrix by definition. This finishes the proof of the lemma. $\square$

## 2.5 Block Orthogonality

Let $\mathbf{x}, \mathbf{y} \in \mathbb{C}^d$ be two nonzero $d$-dimensional vectors. Let $\mathbf{x}'$ and $\mathbf{y}'$ be the two non-negative vectors such that $x_i' = |x_i|$ and $y_i' = |y_i|$ for all $i$, and assume that $\mathbf{x}'$ and $\mathbf{y}'$ are linearly dependent. As a result, these four vectors share the same nonzero entries, and we use $T \subseteq [d]$ to denote the set of such indices. Let

$$\{\mu_1, \ldots, \mu_\ell\} = \{x_i' : i \in T\},$$

for some $\ell \geq 1$, such that $\mu_1 > \cdots > \mu_\ell > 0$. This further partitions $T$ into $T_1, \ldots, T_\ell$ with

$$x_i' = \mu_k, \quad \text{for all } i \in T_k \text{ and } k \in [\ell].$$

It is also clear that $\mathbf{y}'$ would yield the same partition because it is linearly dependent with $\mathbf{x}'$.

Now we say $\mathbf{x}$ and $\mathbf{y}$ are *block-orthogonal* if for every $k \in [\ell]$,

$$\sum_{i \in T_k} x_i \cdot \overline{y_i} = 0. \tag{2}$$

It is also easy to show that $\mathbf{x}$ and $\mathbf{y}$ are *orthogonal* if they are block-orthogonal:

$$\sum_{i \in [d]} x_i \cdot \overline{y_i} = \sum_{i \in T} x_i \cdot \overline{y_i} = \sum_{k \in [\ell]} \sum_{i \in T_k} x_i \cdot \overline{y_i} = 0.$$

We need the following property of two vectors being block-orthogonal:

**Lemma 5.** *If $\mathbf{x}$ and $\mathbf{y}$ are block-orthogonal, and the non-zero entries of these two vectors satisfy $x_i^K > 0$ and $y_i^K > 0$ for some integer $K \geq 1$, then we have*

$$\sum_{i \in D} x_i^{sK+1} \cdot y_i^{rK-1} = 0, \quad \text{for any integers } s \geq 0 \text{ and } r \geq 1.$$

*Proof.* We use the same notation as in the definition of block orthogonality above. For each $i \in T$, we let $z_i = x_i/|x_i|$ and $w_i = y_i/|y_i|$. Then by the assumption of the lemma, $z_i$ and $w_i$ are roots of unity whose order divides $K$. As $\mathbf{x}'$ and $\mathbf{y}'$ are linearly dependent, there are $\nu_1 > \cdots > \nu_\ell > 0$ such that $|y_i| = \nu_k$ for all $i \in T_k$ and $k \in [\ell]$. Now we can rewrite (2) as

$$0 = \sum_{i \in T_k} x_i \cdot \overline{y_i} = \mu_k \cdot \nu_k \sum_{i \in T_k} z_i \cdot \overline{w_i}.$$

Then the lemma follows from

$$\sum_{i \in D} x_i^{sK+1} \cdot y_i^{rK-1} = \sum_{k \in [\ell]} \sum_{i \in T_k} x_i^{sK+1} \cdot y_i^{rK-1} = \sum_{k \in [\ell]} \mu_k^{sK+1} \cdot \nu_k^{rK-1} \sum_{i \in T_k} z_i^{sK+1} \cdot w_i^{rK-1}$$

$$= \sum_{k \in [\ell]} \mu_k^{sK+1} \cdot \nu_k^{rK-1} \sum_{i \in T_k} z_i \cdot \overline{w_i} = 0.$$

The second to the last equation uses the fact that $z_i, w_i$ are roots of unity whose order divides $K$. □

We are now ready to define block-orthogonal functions:

**Definition 1** (Block-Orthogonal Function)**.** *Let $F: D^n \to \mathbb{C}$ be a block-rank-$1$ function with $n \geq 2$. We call it a* block-orthogonal *function if for all $\mathbf{x}, \mathbf{y} \in D^{n-1}$ such that $F(\mathbf{x}, *), F(\mathbf{y}, *) \neq \mathbf{0}$ and $\mathbf{x} \sim_{|F|} \mathbf{y}$, the two vectors $F(\mathbf{x}, *)$ and $F(\mathbf{y}, *)$ are either linearly dependent or block-orthogonal.*

## 2.6 Unweighted Counting CSP

We need the following connection between weighted and *unweighted* #CSP. The latter is the case when all the functions in $\mathcal{F}$ take values in $\{0, 1\}$, for which we adopt the following notation. Let $D = [d]$ be a domain. An unweighted constraint language $\Gamma$ over domain $D$ is a finite set of relations $\{\Phi_1, \Phi_2, \ldots, \Phi_h\}$ in which every $\Phi_i$ is an $r_i$-ary relation over $D^{r_i}$, for some $r_i \geq 1$. $D$ and $\Gamma$ define the following problem which we denote by #CSP($\Gamma$). Let $\mathbf{x} = (x_1, \ldots, x_n) \in D^n$ be a set of $n$ variables over $D$. The input is a collection $I$ of tuples $(\Phi, i_1, \ldots, i_r)$ in which $\Phi$ is an $r$-ary relation in $\Gamma$ and $i_1, \ldots, i_r \in [n]$. The input $I$ then defines the following relation $R_I$ over $D^n$:

$$\mathbf{x} \in R_I \iff \text{for every tuple } (\Phi, i_1, \ldots, i_r) \in I, \text{ we have } (x_{i_1}, \ldots, x_{i_r}) \in \Phi.$$

Given $I$, the output of the problem is the number of $\mathbf{x} \in D^n$ in this relation $R_I$.

Given $F: D^n \to \mathbb{C}$, we use $\Phi_F = \mathsf{Boolean}(F)$ to denote the relation over $n$ variables such that

$$\mathbf{x} \in \Phi_F \iff F(\mathbf{x}) \neq 0, \quad \text{for all } \mathbf{x} \in D^n.$$

The following lemma is a corollary of Lemma 1:

**Lemma 6.** *Given a finite set of complex functions $\mathcal{F} = \{F_1, \ldots, F_h\}$, we have*

$$\text{\#CSP}(\Gamma) \leq_\mathsf{T} \text{\#CSP}(\mathcal{F}),$$

*where $\Gamma = \{\Phi_1, \ldots, \Phi_h\}$ and $\Phi_i = \mathsf{Boolean}(F_i)$ for all $i \in [h]$.*

*Proof.* By Lemma 1, it suffices to show that #CSP($\Gamma$) $\leq_\mathsf{T}$ COUNT($\mathcal{F}$).

Let $I$ be an input instance of #CSP($\Gamma$) over $n$ variables and let $R$ be the relation that it defines. We then construct an instance $I'$ of #CSP($\mathcal{F}$) in polynomial time, by replacing the relation $\Phi_i$ in each tuple of $I$ with its corresponding function $F_i \in \mathcal{F}$, and let $F$ denote the function that $I'$ defines. Then we have

$$\mathbf{x} \in R \iff F(\mathbf{x}) \neq 0, \quad \text{for all } \mathbf{x} \in D^n,$$

and thus,

$$|R| = d^n - \big[\text{the number of } \mathbf{x} \in D^n \text{ such that } F(\mathbf{x}) = 0\big].$$

The right hand side can be obtained by calling a subroutine for COUNT($\mathcal{F}$). □

## 2.7 The Purification Lemma

As it will become clear later, it is easier to work with functions that take complex values with rational arguments. We need the following definition:

**Definition 2** (Pure Functions). *We call $F: D^n \to \mathbb{C}$ a pure complex function if $F(\mathbf{x})$ is the product of a non-negative integer and a root of unity, for all $\mathbf{x} \in D^n$. Given a pure function $F$, we use $\mathsf{order}(F)$ to denote the smallest positive integer $K$ such that $(F(\mathbf{x}))^K$ is positive for all $\mathbf{x} \in D^n$ with $F(\mathbf{x}) \neq 0$.*

A useful tool in proving the hardness part of our dichotomy is the following Purification Lemma. It was introduced in the study of complex graph homomorphisms in [7], and gives us a connection between pure and general functions (which can take values with irrational arguments). In Section 4 and Section 5, we will see two examples where the Purification Lemma is used to extend two hardness lemmas from pure to general functions.

**Lemma 7** (The Purification Lemma). *There is a mapping $\mathsf{Pure}$ which, given any finite tuple $(F_1, \ldots, F_h)$ of complex-valued functions, produces a tuple of pure functions*

$$(F_1', \ldots, F_h') = \mathsf{Pure}(F_1, \ldots, F_h) \tag{3}$$

*in which each $F_i'$ has the same arity $r_i \geq 1$ as $F_i$, such that*

1. *$\#\mathsf{CSP}(F_1', \ldots, F_h') \equiv_\mathsf{T} \#\mathsf{CSP}(F_1, \ldots, F_h)$;*
2. *For every $i \in [h]$, we have $\mathsf{Boolean}(F_i') = \mathsf{Boolean}(F_i)$;*
3. *For every $i \in [h]$ with $r_i \geq 2$, if $F_i'$ is block-rank-1 then $F_i$ is block-rank-1; and*
4. *If $F_i'$ is block-rank-1, then for any $\mathbf{x}, \mathbf{y} \in D^{n-1}$ such that $F_i'(\mathbf{x}, *)$ and $F_i'(\mathbf{y}, *)$ share at least one common nonzero entry, we have*

    (a) *$F_i'(\mathbf{x}, *)$ and $F_i'(\mathbf{y}, *)$ are linearly dependent iff $F_i(\mathbf{x}, *)$ and $F_i(\mathbf{y}, *)$ are linearly dependent;*

    (b) *If $F_i'(\mathbf{x}, *)$ and $F_i'(\mathbf{y}, *)$ are block-orthogonal, then $F_i(\mathbf{x}, *)$ and $F_i(\mathbf{y}, *)$ are block-orthogonal.*

The proof of the Purification Lemma uses the following lemma from [7]. We start with a definition:

**Definition 3.** *Let $C = \{c_1, c_2, \ldots, c_n\}$ be a set of nonzero algebraic numbers, for some $n \geq 1$. Then we say $\{g_1, \ldots, g_s\}$, for some $s \geq 0$, is a generating set of $C$ if*

1. *Every $g_i$ is a nonzero algebraic number in $\mathbb{Q}(C)$;*
2. *For all $(k_1, \ldots, k_s) \in \mathbb{Z}^s - \{\mathbf{0}\}$, $g_1^{k_1} \cdots g_s^{k_s}$ is not a root of unity; and*
3. *For every $c \in C$, there exists a unique tuple $(k_1, \ldots, k_s) \in \mathbb{Z}^s$ such that*

$$\frac{c}{g_1^{k_1} \cdots g_s^{k_s}} \text{ is a root of unity.}$$

Note that $s = 0$ happens if and only if all the $c_i$'s in $C$ are roots of unity.

**Lemma 8** (7.2 in [7]). *Let $C$ be a finite set of nonzero algebraic numbers, then it has a generating set.*

We now use Lemma 8 to prove the Purification Lemma:

*Proof of the Purification Lemma.* We first describe the mapping Pure, and then prove the properties.

By Lemma 8, we use $\{g_1, \ldots, g_s\}$ to denote a generating set of $\mathsf{Im}(F_1, \ldots, F_h) - \{0\}$. Given a tuple $\mathbf{k} = (k_1, \ldots, k_s) \in \mathbb{Z}^s$, we use $g(\mathbf{k})$ to denote $g_1^{k_1} \cdots g_s^{k_s}$ for convenience. By definition, there is a unique tuple $\mathbf{k} \in \mathbb{Z}^s$ for each $c \in \mathsf{Im}(F_1, \ldots, F_h) - \{0\}$ such that $c/g(\mathbf{k})$ is a root of unity. Because the functions $F_1, \ldots, F_h$ are fixed, all the integers in $\mathbf{k}$ are considered as constants.

We define $F_i'$ from $F_i$ as follows. For every $\mathbf{x} \in D^{r_i}$, $F_i'(\mathbf{x}) = 0$ if $F_i(\mathbf{x}) = 0$. If $F_i(\mathbf{x}) \neq 0$, then there exists a unique tuple $\mathbf{k} \in \mathbb{Z}^s$ such that $F_i(\mathbf{x})/g(\mathbf{k})$ is a root of unity, and we set

$$F_i'(\mathbf{x}) = p_1^{k_1} \cdots p_s^{k_s} \cdot \frac{F_i(\mathbf{x})}{g(\mathbf{k})},$$

where $p_i$ denotes the $i^{\text{th}}$ smallest prime. It is clear that Property 2 of the lemma is satisfied. In the rest of the proof, we will use $p(\mathbf{k})$ to denote $p_1^{k_1} \cdots p_s^{k_s}$ for all $\mathbf{k} \in \mathbb{Z}^s$.

Next we prove the equivalence of the two problems. By Lemma 1, it suffices to show that

$$\mathsf{COUNT}(F_1, \ldots, F_h) \equiv_T \mathsf{COUNT}(F_1', \ldots, F_h'). \tag{4}$$

We start with the reduction from $\mathsf{COUNT}(F_1, \ldots, F_h)$ to $\mathsf{COUNT}(F_1', \ldots, F_h')$.

Given an instance $I$ of $\#\mathsf{CSP}(F_1, \ldots, F_h)$ over $n$ variables, we use $I'$ to denote the instance of $\#\mathsf{CSP}(F_1', \ldots, F_h')$ obtained by replacing the $F_i$ in each tuple of $I$ with its corresponding function $F_i'$. Also let $m = |I| = |I'|$ and let $F$ and $F'$ denote the functions that $I$ and $I'$ define, respectively. By Property 2, we have $F(\mathbf{x}) \neq 0$ iff $F'(\mathbf{x}) \neq 0$ and thus, the number of $\mathbf{x}$ such that $F(\mathbf{x}) = 0$ is the same as the number of $\mathbf{x}$ such that $F'(\mathbf{x}) = 0$. The latter can be obtained by calling a subroutine for $\mathsf{COUNT}(F_1', \ldots, F_h')$.

Let $\{c_1, \ldots, c_t\} = \mathsf{Im}(F_1, \ldots, F_h) - \{0\}$ with $t$ being a constant since the set of functions is fixed. We then compute the following set $C_m$ in time polynomial in $m$:

$$C_m = \left\{ c_1^{\ell_1} \cdots c_t^{\ell_t} : \ell_1, \ldots, \ell_t \text{ are non-negative integers and } \ell_1 + \cdots + \ell_t = m \right\}.$$

For each $c \in C_m$ we also compute the unique tuple $\mathbf{k} \in \mathbb{Z}^s$ such that $c/g(\mathbf{k})$ is a root of unity, using the known tuples for $\{c_1, \ldots, c_t\}$. By the construction of $F_1', \ldots, F_h'$ from $F_1, \ldots, F_h$, and by the assumption that $\{g_1, \ldots, g_s\}$ is a generating set, we have

$$F(\mathbf{x}) = c \iff F'(\mathbf{x}) = p(\mathbf{k}) \cdot \frac{c}{g(\mathbf{k})}, \quad \text{for all } \mathbf{x} \in D^n.$$

Hence, the number of $\mathbf{x}$ with $F(\mathbf{x}) = c$ can be obtained by calling a subroutine for $\mathsf{COUNT}(F_1', \ldots, F_h')$. The other direction from $\mathsf{COUNT}(F_1', \ldots, F_h')$ to $\mathsf{COUNT}(F_1, \ldots, F_h)$ can be proved similarly.

Now we check Property 3. In the rest of the proof, we use $F$ to denote $F_i$, $F'$ to denote $F_i'$ and $r$ to denote $r_i$, the arity of $F_i$, for convenience.

Assume $r \geq 2$ and $F'$ is block-rank-1. Let $\mathbf{x}, \mathbf{y} \in D^{r-1}$ be two vectors such that $F(\mathbf{x}, *)$ and $F(\mathbf{y}, *)$ share at least one common nonzero entry. From Property 2 and the assumption that $F'$ is block-rank-1 we know that $|F'(\mathbf{x}, *)|$ and $|F'(\mathbf{y}, *)|$ must be nonzero and linearly dependent.

To prove that $|F(\mathbf{x}, *)|$ and $|F(\mathbf{y}, *)|$ are linearly dependent, it suffices to show for all indices $i, j \in D$

of nonzero entries of $F(\mathbf{x}, *)$ (which are also indices of nonzero entries of $F(\mathbf{y}, *), F'(\mathbf{x}, *), F'(\mathbf{y}, *)$),

$$|F(\mathbf{x}, i)| \cdot |F(\mathbf{y}, j)| = |F(\mathbf{y}, i)| \cdot |F(\mathbf{x}, j)|. \tag{5}$$

To this end, we let $\mathbf{u}, \mathbf{v}, \mathbf{w}, \mathbf{z} \in \mathbb{Z}^s$ denote the vectors such that

$$\frac{F'(\mathbf{x}, i)}{p(\mathbf{u})}, \quad \frac{F'(\mathbf{y}, j)}{p(\mathbf{v})}, \quad \frac{F'(\mathbf{y}, i)}{p(\mathbf{w})}, \quad \frac{F'(\mathbf{x}, j)}{p(\mathbf{z})}$$

are all roots of unity. Because $|F'(\mathbf{x}, *)|$ and $|F'(\mathbf{y}, *)|$ are linearly dependent, we have

$$p_1^{u_1+v_1} \cdots p_s^{u_s+v_s} = |F'(\mathbf{x}, i)| \cdot |F'(\mathbf{y}, j)| = |F'(\mathbf{y}, i)| \cdot |F'(\mathbf{x}, j)| = p_1^{w_1+z_1} \cdots p_s^{w_s+z_s}$$

and thus, $u_k + v_k = w_k + z_k$ for all $k \in [s]$. (5) then follows directly from the construction of $F'$.

Next we prove Property 4(a). Assume that $F(\mathbf{x}, *)$ and $F(\mathbf{y}, *)$ are linearly dependent. Then we use $i, j \in D$ to denote two indices of nonzero entries of $F'(\mathbf{x}, *)$, which must be indices of nonzero entries of $F'(\mathbf{y}, *), F(\mathbf{x}, *)$ and $F(\mathbf{y}, *)$ as well. Similarly, let $\mathbf{u}, \mathbf{v}, \mathbf{w}, \mathbf{z} \in \mathbb{Z}^s$ be the vectors such that

$$c_1 = \frac{F(\mathbf{x}, i)}{g(\mathbf{u})}, \quad c_2 = \frac{F(\mathbf{y}, j)}{g(\mathbf{v})}, \quad c_3 = \frac{F(\mathbf{y}, i)}{g(\mathbf{w})}, \quad c_4 = \frac{F(\mathbf{x}, j)}{g(\mathbf{z})},$$

and $c_1, \ldots, c_4$ are all roots of unity. Because $F(\mathbf{x}, *)$ and $F(\mathbf{y}, *)$ are linearly dependent, we have

$$c_1 \cdot c_2 \cdot g_1^{u_1+v_1} \cdots g_s^{u_s+v_s} = F(\mathbf{x}, i) \cdot F(\mathbf{y}, j) = F(\mathbf{y}, i) \cdot F(\mathbf{x}, j) = c_3 \cdot c_4 \cdot g_1^{w_1+z_1} \cdots g_s^{w_s+z_s}.$$

By the definition of generating sets, we must have $c_1 \cdot c_2 = c_3 \cdot c_4$ and $u_k + v_k = w_k + z_k$ for all $k \in [s]$. On the other hand, by the construction of $F'$, we have

$$F'(\mathbf{x}, i) \cdot F'(\mathbf{y}, j) = c_1 \cdot c_2 \cdot p_1^{u_1+v_1} \cdots p_s^{u_s+v_s} = c_3 \cdot c_4 \cdot p_1^{w_1+z_1} \cdots p_s^{w_s+z_s} = F'(\mathbf{y}, i) \cdot F'(\mathbf{x}, j)$$

and thus, $F'(\mathbf{x}, *)$ and $F'(\mathbf{y}, *)$ are also linearly dependent. The other direction can be proved similarly.

For 4(b), assume that $F'(\mathbf{x}, *)$ and $F'(\mathbf{y}, *)$ are block-orthogonal. Note that when $F'$ is block-rank-1 $F$ is also block-rank-1 by Property 3. We then use $T \subseteq D$ to denote the set of indices $j \in D$ such that $F(\mathbf{x}, j) \neq 0$ (and $F(\mathbf{y}, j), F'(\mathbf{x}, j), F'(\mathbf{y}, j) \neq 0$ as both functions are block-rank-1). We use $F'(\mathbf{x}, *)$ to further partition $T$ into $T_1, \ldots, T_t$ for some $t \geq 1$: there are positive integers $\mu_1 > \cdots > \mu_t > 0$ such that

$$|F'(\mathbf{x}, j)| = \mu_k, \quad \text{for all } j \in T_k \text{ and } k \in [t].$$

Since $F'$ is block-rank-1, we know $|F'(\mathbf{x}, *)|, |F'(\mathbf{y}, *)|$ are linearly dependent and thus, there are positive integers $\omega_1 > \cdots > \omega_t > 0$ such that $(\mu_1, \ldots, \mu_t)$ and $(\omega_1, \ldots, \omega_t)$ are linearly dependent and

$$|F'(\mathbf{y}, j)| = \omega_k, \quad \text{for all } j \in T_k \text{ and } k \in [t].$$

We also use $c(\mathbf{x}, j)$ and $c(\mathbf{y}, j)$ to denote the roots of unity such that

$$F'(\mathbf{x}, j) = \mu_k \cdot c(\mathbf{x}, j) \quad \text{and} \quad F'(\mathbf{y}, j) = \omega_k \cdot c(\mathbf{y}, j).$$

Because $F'(\mathbf{x}, *)$ and $F'(\mathbf{y}, *)$ are block-orthogonal, by definition, we have

$$\sum_{j \in T_k} F'(\mathbf{x}, j) \cdot \overline{F'(\mathbf{y}, j)} = \mu_k \cdot \omega_k \sum_{j \in T_k} c(\mathbf{x}, j) \cdot \overline{c(\mathbf{y}, j)} = 0, \quad \text{for all } k \in [t]. \tag{6}$$

For each $k \in [t]$, we use $\mathbf{u}_k$ and $\mathbf{v}_k \in \mathbb{Z}^s$ to denote the two unique vectors such that

$$\mu_k = p(\mathbf{u}_k) \quad \text{and} \quad \omega_k = p(\mathbf{v}_k).$$

Then by the construction of $F'$ from $F$, we have for all $j \in T_k$,

$$\begin{aligned} F(\mathbf{x}, j) &= g(\mathbf{u}_k) \cdot c(\mathbf{x}, j) \quad \text{and} \quad |F(\mathbf{x}, j)| = |g(\mathbf{u}_k)| \\ F(\mathbf{y}, j) &= g(\mathbf{v}_k) \cdot c(\mathbf{y}, j) \quad \text{and} \quad |F(\mathbf{y}, j)| = |g(\mathbf{v}_k)|. \end{aligned} \tag{7}$$

Now we are ready to prove that $F(\mathbf{x}, *)$ and $F(\mathbf{y}, *)$ are block-orthogonal. Let $v = |F(\mathbf{x}, j)| > 0$ for some $j \in T$ and let $S_v \subseteq T$ be the set of indices $j$ such that $|F(\mathbf{x}, j)| = v$. Then by (7) above $S_v$ must be the union of some of the $T_k$'s. Without loss of generality, let $S_v = T_1 \cup \cdots \cup T_q$ for some $q \leq t$, then

$$\sum_{j \in S_v} F(\mathbf{x}, j) \cdot \overline{F(\mathbf{y}, j)} = \sum_{k \in [q]} \sum_{j \in T_k} g(\mathbf{u}_k) c(\mathbf{x}, j) \cdot \overline{g(\mathbf{v}_k) c(\mathbf{y}, j)} = \sum_{k \in [q]} g(\mathbf{u}_k) \overline{g(\mathbf{v}_k)} \sum_{j \in T_k} c(\mathbf{x}, j) \cdot \overline{c(\mathbf{y}, j)} = 0$$

where the first equation uses (7) and the last equation uses (6). This finishes the proof. □

We remark that in both Property 3 and Property 4(b) of the lemma, the statement only holds in one direction. For example, when $F_i$ is block-rank-1, it is not clear how to prove that $F'_i$ is block-rank-1 as well. However, it turns out that the directions that we can prove are the ones that we will actually need later in proving those hardness lemmas for general functions.

Using Property 2, 3 and 4 of the Purification Lemma, we have the following corollary:

**Corollary 1.** *Given (3), if $F'_i$ is block-orthogonal, then so is $F_i$. Moreover, the equivalence relations $\sim_{F_i}$ and $\sim_{F'_i}$, as defined by $F_i$ and $F'_i$, are the same.*

### 2.8 Mal'tsev Polymorphisms and Witness Functions

The algorithmic part of our dichotomy theorem uses the following concept of *Mal'tsev polymorphisms*:

**Definition 4.** *Let $\Phi \subseteq D^n$ be an $n$-ary relation and $\varphi : D^3 \to D$ be a map. If for any $\mathbf{u}, \mathbf{v}, \mathbf{w} \in \Phi$,*

$$\big(\varphi(u_1, v_1, w_1), \ldots, \varphi(u_n, v_n, w_n)\big) \in \Phi,$$

*then we say $\Phi$ is closed under $\varphi$, and call $\varphi$ a ternary polymorphism of $\Phi$.*

*Given $\Phi \subseteq D^n$ and $\varphi : D^3 \to D$, we use $\mathsf{cl}_\varphi \Phi$ to denote the closure of $\Phi$ under $\varphi$, that is, the smallest relation that contains $\Phi$ and is closed under $\varphi$.*

**Definition 5** (Mal'tsev Polymorphism). *Let $\Phi \subseteq D^n$ be an $n$-ary relation, then we say $\varphi : D^3 \to D$ is a Mal'tsev polymorphism of $\Phi$ if $\varphi$ is a polymorphism of $\Phi$ and satisfies*

$$\varphi(a, b, b) = \varphi(b, b, a) = a, \quad \text{for all } a, b \in D. \tag{8}$$

Let $\Gamma = \{\Phi_1, \ldots, \Phi_h\}$ be a finite set of relations, then we say $\varphi$ is a Mal'tsev polymorphism *of $\Gamma$ if it is a Mal'tsev polymorphism of $\Phi_i$ for all $i \in [h]$.*

Bulatov and Dalmau [2] gave the following #P-hardness theorem. Also see Dyer and Richerby [13].

**Theorem 3** ([2]). *Let $\Gamma = \{\Phi_1, \ldots, \Phi_h\}$ be a finite set of relations. If $\Gamma$ does not have any Mal'tsev polymorphism, then $\#\mathsf{CSP}(\Gamma)$ is #P-hard.*

**Corollary 2.** *Let $\Lambda$ be a collection of (possibly infinitely many) relations. Either all relations in $\Lambda$ share a common Mal'tsev polymorphism $\varphi$; or there is a finite subset $\Gamma \subset \Lambda$ such that $\#\mathsf{CSP}(\Gamma)$ is #P-hard.*

*Proof.* Note that, given $d$, there are only finitely many maps $\varphi: D^3 \to D$. We let $P$ denote the set of all such maps. Now assume the relations in $\Lambda$ do not share a common Mal'tsev polymorphism, then for any $\varphi \in P$, there is a relation $\Phi_\varphi \in \Lambda$ of which $\varphi$ is not a Mal'tsev polymorphism. Then from Theorem 3, we know that $\#\mathsf{CSP}(\Gamma)$ is #P-hard, where $\Gamma = \{\Phi_\varphi : \varphi \in P\}$ is a finite subset of $\Lambda$. □

Let $\Phi$ be an $n$-ary relation with variables $x_1, \ldots, x_n$ ranging over $D$, then in general $|\Phi|$ could be exponentially large in $n$. But when $\Phi$ is known to have a Mal'tsev polymorphism and such a polymorphism $\varphi$ is also given, Dyer and Richerby introduced in [13] the following succinct representation for $\Phi$, which is similar to the "compact representation" of Bulatov and Dalmau [3]. We start with some notation.

For each $i \in [n]$ we define the following relation $\sim_i$ on $\mathsf{Pr}_i \Phi$, the projection of $\Phi$ on its $i^{\text{th}}$ coordinate: $a \sim_i b$ if there exist tuples $\mathbf{x} \in D^{i-1}$ and $\mathbf{y}_a, \mathbf{y}_b \in D^{n-i}$ such that

$$\mathbf{x} \circ a \circ \mathbf{y}_a \in \Phi \quad \text{and} \quad \mathbf{x} \circ b \circ \mathbf{y}_b \in \Phi.$$

For the case when $i = 1$, we have $a \sim_1 b$ for all $a, b \in \mathsf{Pr}_1 \Phi$ because they share the common empty prefix $\epsilon$. It was then shown in [13] that if $\Phi$ has a Mal'tsev polymorphism, $\sim_i$ must be an equivalence relation:

**Lemma 9.** *If $\Phi$ has a Mal'tsev polymorphism, then $\sim_i$ is an equivalence relation for all $i \in [n]$.*

When $\Phi$ has a Mal'tsev polymorphism, we let $\mathcal{E}_{i,k} \subseteq \mathsf{Pr}_i \Phi$, where $k = 1, 2, \ldots$, denote the equivalence classes of $\sim_i$. Moreover, we can use the Mal'tsev polymorphism to show that

**Lemma 10.** *If $a \sim_i b$ and $\mathbf{x} \in \Phi$ with $x_i = a$, then there is a $\mathbf{y} \in \Phi$ with $y_i = b$ and $\mathsf{Pr}_{[i-1]} \mathbf{x} = \mathsf{Pr}_{[i-1]} \mathbf{y}$.*

*Proof.* Because $a \sim_i b$, by definition, there exist $\mathbf{z} \in D^{i-1}$ and $\mathbf{u}_1, \mathbf{u}_2 \in D^{n-i}$ such that

$$\mathbf{z} \circ a \circ \mathbf{u}_1 \in \Phi \quad \text{and} \quad \mathbf{z} \circ b \circ \mathbf{u}_2 \in \Phi.$$

Applying a Mal'tsev polymorphism $\varphi$ of $\Phi$ on these two vectors together with $\mathbf{x} \in \Phi$ gives a new vector $\mathbf{y} \in \Phi$. It is easy to check that $\mathbf{y}$ satisfies both properties, and the lemma is proven. □

Next, we define the succinct representation called witness functions [13].

**Definition 6** (Witness Function). *Let $\Phi \subseteq D^n$ be a relation that has a Mal'tsev polymorphism, then we say $\omega: [n] \times D \to D^n \cup \{\bot\}$ is a* witness function *of $\Phi$ if*

1. *For any $i \in [n]$ and $a \notin \mathsf{Pr}_i \Phi$, $\omega(i, a) = \bot$;*
2. *For any $i \in [n]$ and $a \in \mathsf{Pr}_i \Phi$, $\omega(i, a) \in \Phi$ is a witness for $(i, a)$, i.e., its ith component is $a$;*

3. For any $i \in [n]$ and $a, b \in \mathsf{Pr}_i \Phi$ with $a \sim_i b$, we have
$$\mathsf{Pr}_{[i-1]} \omega(i, a) = \mathsf{Pr}_{[i-1]} \omega(i, b).$$

In [13] a subset of $\Phi$ that contains the image of a witness function of $\Phi$ is called a *frame* of $\Phi$. But in this paper, we will only use witness functions. The following lemma from [13] is the reason why a witness function is considered as a succinct (and linear-size) representation of $\Phi$:

**Lemma 11** (Membership). *Let $\Phi \subseteq D^n$ be an $n$-ary relation which has a Mal'tsev polymorphism. With $\omega$, a witness function of $\Phi$, and $\varphi$, a Mal'tsev polymorphism of $\Phi$, we can solve the following problem in time polynomial in $n$: Given $\mathbf{x} \in D^n$, decide if $\mathbf{x} \in \Phi$ or not.*

It is also easy to show that, if $\varphi$ is a Mal'tsev polymorphism of $\Phi \subseteq D^n$, then all three operations on $\Phi$ described in Section 2.1, that is, projection, pinning, and permutation, would result in a relation $\Phi'$ of which $\varphi$ remains a Mal'tsev polymorphism.

**Lemma 12.** *Let $\varphi$ be a Mal'tsev polymorphism of $\Phi \subseteq D^n$. Let $\ell \in [n]$, $\mathbf{a} \in D^\ell$, and $\pi$ be a permutation on $[n]$. Then $\varphi$ is also a Mal'tsev polymorphism of $\mathsf{Pr}_{[\ell]} \Phi$, $\Phi(\mathbf{a}, *)$, and $\Phi_\pi$.*

Moreover, given a witness function $\omega$ of $\Phi$, we can construct a witness function of $\mathsf{Pr}_{[\ell]} \Phi$, $\Phi(\mathbf{a}, *)$ and $\Phi_\pi$ in time polynomial in $n$. For pinning and projection, the following two lemmas can be found in Dyer and Richerby [13]. In Section 7, we will discuss permutation and two other polynomial-time operations on witness functions, *union* and *splitting*. They will play an important role in the algorithmic part.

**Lemma 13** (Pinning). *Let $\omega : [n] \times D \to D^n \cup \{\bot\}$ be a witness function of $\Phi$. Then given any $\mathbf{a} \in D^\ell$ for some $\ell \in [n]$, we can construct a witness function for $\Phi(\mathbf{a}, *)$ in time polynomial in $n$.*

**Lemma 14** (Projection). *Let $\omega : [n] \times D \to D^n \cup \{\bot\}$ be a witness function of $\Phi$. Given any $\ell \in [n]$, we can construct a witness function for $\mathsf{Pr}_{[\ell]} \Phi$ in time polynomial in $n$.*

*When $\ell$ is bounded by a constant, we can use $\omega$ to compute the projection $\mathsf{Pr}_{[\ell]} \Phi$ itself in polynomial time. Given an $\mathbf{x} \in \mathsf{Pr}_{[\ell]} \Phi$, we can also compute a vector $\mathbf{y} \in \Phi$ with $\mathbf{x} = \mathsf{Pr}_{[\ell]} \mathbf{y}$ in polynomial time.*

Let $\Gamma = \{\Phi_1, \ldots, \Phi_h\}$ be an unweighted language over $D$. Then by Theorem 3, $\#\mathsf{CSP}(\Gamma)$ is $\#\mathsf{P}$-hard if the relations in $\Gamma$ do not share a common Mal'tsev polymorphism. Dyer and Richerby showed [13] that if all the relations in $\Gamma$ share a common Mal'tsev polymorphism, then given any instance $I$ of $\#\mathsf{CSP}(\Gamma)$, a witness function for the relation $R_I$ that $I$ defines can be constructed efficiently.

**Theorem 4.** *Let $\varphi$ be a Mal'tsev polymorphism of all the relations in $\Gamma$, then given any input instance $I$ of $\#\mathsf{CSP}(\Gamma)$, one can construct a witness function of $R_I$ in polynomial time.*

### 2.9 Type-Partition Maps

Finally, we define type-partition maps. Let $S \subseteq D^n$ be a nonempty set and let $S_1, \ldots, S_k$ be a partition of $S$, for some $k \geq 1$: the $S_i$'s are nonempty and pairwise disjoint, and $S = S_1 \cup \cdots \cup S_k$. Then the pair $(S, (S_1, \ldots, S_k))$ defines the following type map $\mathsf{type}(\cdot)$: Given any $\ell \in [n]$ and $\mathbf{x} \in D^\ell$,

$$\mathsf{type}(\mathbf{x}) = \Big\{ j \in [k] : \exists \mathbf{y} \in S_j \text{ such that } \mathbf{x} = \mathsf{Pr}_{[\ell]} \mathbf{y} \Big\} \subseteq [k]. \tag{9}$$

We usually call type(**x**) the type of **x**. When $\ell = n$, type(**x**) is either the empty set or a singleton. When $\ell = n$ and type(**x**) is a singleton, we sometimes also refer to the element in type(**x**) simply as the type of **x** for convenience.

**Definition 7.** *Let* type(**x**) *be the map defined by* $(S, (S_1, \ldots, S_k))$, *then we say it is a* type-partition map *if for all* $\ell \in [n]$ *and* $\mathbf{x}, \mathbf{y} \in D^\ell$, type(**x**) *and* type(**y**) *are either the same or disjoint.*

Now let type($\cdot$), as defined by $(S, (S_1, \ldots, S_k))$ in (9), be a type-partition map over $D^n$, then we will refer to the following $(n+1)$-tuple

$$\mathfrak{T} = (\mathcal{T}_0, \mathcal{T}_1, \ldots, \mathcal{T}_n), \quad \text{where } \mathcal{T}_\ell = \left\{ \text{type}(\mathbf{x}) \subseteq [k] : \mathbf{x} \in D^\ell \text{ and type}(\mathbf{x}) \neq \emptyset \right\} \subset 2^{[k]}.$$

as the *list of types* of type($\cdot$). Here for the special case of $\ell = 0$, we have $\mathcal{T}_0 = \{[k]\}$.

Because it is assumed to be a type-partition map, we have $|\mathcal{T}_\ell| \leq k$ for all $\ell$. It is also clear, from the definition, that all the sets in $\mathcal{T}_\ell$ are nonempty, since we are only interested in $\mathbf{x} \in D^\ell$ with type(**x**) $\neq \emptyset$. It is easy to prove the following lemma:

**Lemma 15.** *Let* type($\cdot$) *be a type-partition map. Then for any* $\ell : 0 \leq \ell \leq n$,

$$\bigcup_{T \in \mathcal{T}_\ell} T = [k].$$

*For any* $i, j : 0 \leq i < j \leq n$ *and* $U \in \mathcal{T}_i, V \in \mathcal{T}_j$, *either* $V \subseteq U$ *or* $U \cap V = \emptyset$.

One way to better understand the list $\mathfrak{T}$ is to consider it as a tree of height $n$: $[k] \in \mathcal{T}_0$ is the root, and the sets of $\mathcal{T}_\ell$ are nodes at level $\ell$ of the tree; $U \in \mathcal{T}_\ell$ and $V \in \mathcal{T}_{\ell+1}$ are connected if $V \subseteq U$. The tree has the property that the leaves are singletons and every other node is the union of its children.

## 3 A Complexity Dichotomy for #CSP with Complex Weights

We now prove Theorem 1. We start by describing the three necessary conditions for tractability.

Let $D = [d]$ be a domain and let $\mathcal{F}$ be a finite set of algebraic complex functions over $D$. We use $\mathcal{W}_\mathcal{F}$ to denote the following set of infinitely many (though countable) complex-valued functions:

$$\mathcal{W}_\mathcal{F} = \left\{ F^{[t]} : F \text{ is a function defined by an input instance of } \#\mathsf{CSP}(\mathcal{F}) \text{ and } t : 1 \leq t \leq \text{arity of } F \right\}.$$

The following lemma concerning $\mathcal{W}_\mathcal{F}$ is easy to prove:

**Lemma 16.** *For any finite subset* $\mathcal{F}' \subset \mathcal{W}_\mathcal{F}$, *we have* $\#\mathsf{CSP}(\mathcal{F}') \leq_\mathsf{T} \#\mathsf{CSP}(\mathcal{F})$.

### 3.1 Hardness Part of the Dichotomy

The hardness part of the dichotomy theorem consists of three conditions over $\mathcal{W}_\mathcal{F}$. The violation of any of these conditions implies that $\#\mathsf{CSP}(\mathcal{F})$ is #P-hard. First, we impose the following condition:

**Block Orthogonality**: Let $\{F_1, \ldots, F_k\}$ be any finite subset of $\mathcal{W}_\mathcal{F}$, and let

$$(F'_1, \ldots, F'_k) = \mathsf{Pure}(F_1, \ldots, F_k).$$

Then for every $F'_i$ with arity $\geq 2$, we have $F'_i$ is *block-orthogonal* (and in particular, *block-rank*-1).

We prove the following lemma in Section 4:

**Lemma 17.** *If $\mathcal{F}$ does not satisfy the* Block Orthogonality *condition, then* $\#\mathsf{CSP}(\mathcal{F})$ *is* $\#$P-*hard.*

Assume $\mathcal{F}$ satisfies the Block Orthogonality condition. By Corollary 1, every $F$ in $\mathcal{W}_\mathcal{F}$ with arity $\geq 2$ is block-orthogonal (and in particular, block-rank-1). Let $n \geq 2$ be the arity of $F \in \mathcal{W}_\mathcal{F}$, and let

$$\{(S_1, \mathbf{v}_1), \ldots, (S_k, \mathbf{v}_k)\}, \quad \text{where } S_j \subseteq D^{n-1} \text{ and } \mathbf{v}_1, \ldots, \mathbf{v}_k \text{ are linearly independent vectors,} \quad (10)$$

denote the row representation of $F$. We have $k \leq d$ because $F$ is block-orthogonal. Let

$$\Psi_F = S_1 \cup \cdots \cup S_k,$$

and let $\mathsf{type}_F(\cdot)$ denote the type map defined by the pair $(\Psi_F, (S_1, \ldots, S_k))$.

Assuming $\mathcal{F}$ satisfies the Block Orthogonality condition. Here is the second condition on $\mathcal{W}_\mathcal{F}$:

**Type Partition**: For any $F \in \mathcal{W}_\mathcal{F}$ of arity $n \geq 2$, $\mathsf{type}_F(\cdot)$ is a type-partition map.

We prove the following hardness lemma in Section 5.

**Lemma 18.** *If $\mathcal{F}$ does not satisfy the* Type Partition *condition, then* $\#\mathsf{CSP}(\mathcal{F})$ *is* $\#$P-*hard.*

We also need a condition on relations defined from $\mathcal{W}_\mathcal{F}$. Assume $\mathcal{F}$ satisfies the Block Orthogonality condition. Let $F \in \mathcal{W}_\mathcal{F}$, and $\Phi_F = \mathsf{Boolean}(F)$. If $F$ has arity $n \geq 2$, we denote its row representation by (10) and define the following relation $\Omega_F$ on $2(n-1)$ variables $\mathbf{x} = (x_1, \ldots, x_{n-1})$ and $\mathbf{y} = (y_1, \ldots, y_{n-1})$:

$$(\mathbf{x}, \mathbf{y}) \in \Omega_F \iff \mathbf{x}, \mathbf{y} \in S_j \text{ for some } j \in [k] \iff F(\mathbf{x}, *), F(\mathbf{y}, *) \text{ are nonzero and linearly dependent}$$

This gives us the following set $\Lambda_\mathcal{F}$ of infinitely many relations derived from the functions in $\mathcal{W}_\mathcal{F}$:

$$\Lambda_\mathcal{F} = \left\{\Phi_F : F \in \mathcal{W}_\mathcal{F}\right\} \cup \left\{\Omega_F : F \in \mathcal{W}_\mathcal{F} \text{ of arity} \geq 2\right\}.$$

We now impose the last condition on $\Lambda_\mathcal{F}$:

**Mal'tsev**: All relations in $\Lambda_\mathcal{F}$ share a common Mal'tsev polymorphism $\varphi : D^3 \to D$.

To finish the hardness part, we prove the following hardness lemma in Section 6:

**Lemma 19.** *If $\mathcal{F}$ does not satisfy the* Mal'tsev *condition, then* $\#\mathsf{CSP}(\mathcal{F})$ *is* $\#$P-*hard.*

### 3.2 Algorithmic Part of the Dichotomy

We show that if a finite set $\mathcal{F}$ of complex functions satisfies all three conditions:

(a) the Block Orthogonality condition
(b) the Type Partition condition

(c) the Mal'tsev condition

then there is a polynomial-time algorithm for $\#\mathsf{CSP}(\mathcal{F})$. Theorem 1 then follows.

First, by the Mal'tsev condition, all the relations in $\Lambda_\mathcal{F}$ share a common Mal'tsev polymorphism. We may assume that such a polymorphism $\varphi$ is given, and will use it later in the algorithm.

Let $I$ be an instance of $\#\mathsf{CSP}(\mathcal{F})$, and let $F : D^n \to \mathbb{C}$ be the function it defines. To compute $Z(F)$, we examine the functions $F = F^{[n]}, \ldots, F^{[2]}$. For each $F^{[t]}$, $t : 2 \le t \le n$, we use

$$\mathcal{S}^{[t]} = \left\{ \left( S^{[t,j]}, \mathbf{v}^{[t,j]} \right) : j \in [s_t] \right\} \tag{11}$$

to denote the row representation of $F^{[t]}$. At this moment, we do not know what exactly $s_t$ is, though by the Block Orthogonality condition, we know $F^{[t]} \in \mathcal{W}_\mathcal{F}$ is block-orthogonal and thus, $0 \le s_t \le d$.

Next by using the Mal'tsev condition, we know that $\varphi$ is a Mal'tsev polymorphism of $\Omega_{F^{[t]}}$, a relation over $2(t-1)$ variables. The following lemma shows that $\varphi$ must also be a Mal'tsev polymorphism of the $S^{[t,j]}$'s when viewed as relations over $t-1$ variables:

**Lemma 20.** *If $\varphi$ is a Mal'tsev polymorphism of $\Omega_{F^{[t]}}$, then it is a Mal'tsev polymorphism of the $S^{[t,j]}$'s.*

*Proof.* Let $\mathbf{u} \in D^{t-1}$ be a vector in $S^{[t,j]}$ (since $S^{[t,j]}$ is nonempty). By the definition of $\Omega_{F^{[t]}}$, we have

$$\mathbf{y} \in S^{[t,j]} \iff (\mathbf{u}, \mathbf{y}) \in \Omega_{F^{[t]}}$$

As a result, $S^{[t,j]} = \Omega_{F^{[t]}}(\mathbf{u}, *)$, and the lemma follows directly from Lemma 13. $\square$

Now it makes sense to talk about witness functions for the $S^{[t,j]}$'s (even though we still do not know $s_t$ at this moment). We prove the following important algorithmic lemma in Section 7 and 8:

**Lemma 21.** *Assume $\mathcal{F}$ satisfies all three conditions with $\varphi$ being a Mal'tsev polymorphism of $\Lambda_\mathcal{F}$. Given an instance $I$ of $\#\mathsf{CSP}(\mathcal{F})$, letting $F : D^n \to \mathbb{C}$ denote the function it defines, we can compute in polynomial time a sequence of $n - 1$ non-negative integers $s_n, \ldots, s_2 \le d$ such that $s_t$ is the number of pairs in the row representation of $F^{[t]}$. Moreover, we can compute in polynomial time $s_t$ pairs for each $2 \le t \le d$:*

$$\left\{ \left( \omega^{[t,j]}, \mathbf{v}^{[t,j]} \right) : j \in [s_t] \right\} \tag{12}$$

*where $\omega^{[t,j]} : [t-1] \times D \to D^{t-1} \cup \{\bot\}$ and $\mathbf{v}^{[t,j]}$ is a nonzero $d$-dimensional vector, such that*

1. $\left\{ \mathbf{v}^{[t,j]} : j \in [s_t] \right\}$ *are exactly the $s_t$ vectors in the row representation of $F^{[t]}$; and*

2. $\omega^{[t,j]}$ *is a witness function of $S^{[t,j]}$, the set paired with $\mathbf{v}^{[t,j]}$ in the row representation of $F^{[t]}$.*

Once we have obtained $s_t$ and the pairs in (12) for each $t$, $Z(F)$ can be computed efficiently:

**Lemma 22** (Computation of $Z(F)$). *Given $s_t$ and (12), $Z(F)$ can be computed in polynomial time.*

*Proof.* For each $t$, we let (11) denote the row representation of $F^{[t]}$. By Lemma 21, all the vectors $\mathbf{v}^{[t,j]}$ in (11) have been computed and for each set $S^{[t,j]}$, we have computed one of its witness functions $\omega^{[t,j]}$.

For any $a_1 \in D$, we will show how to compute $F^{[1]}(a_1)$ efficiently. The lemma then follows because

$$Z(F) = \sum_{a_1 \in D} F^{[1]}(a_1).$$

We start with an informal description of the algorithm.

We first check whether $a_1 \in S^{[2,j]}$ for some $j \in [s_2]$. This can be done efficiently as $s_2 \leq d$ is bounded by a constant and for each $j \in [s_2]$, whether $a_1 \in S^{[2,j]}$ or not can be checked efficiently using the witness function $\omega^{[2,j]}$ of $S^{[2,j]}$. By definition, if $a_1 \notin S^{[2,j]}$ for all $j \in [s_t]$, we must have $F^{[2]}(a_1, *) = \mathbf{0}$ and thus,

$$F^{[1]}(a_1) = \sum_{b \in D} F^{[2]}(a_1, b) = 0.$$

Otherwise, let $j \in [s_t]$ be the unique index such that $a_1 \in S^{[2,j]}$, and $a_2 \in D$ be the smallest nonzero index of $\mathbf{v}^{[2,j]}$. By the definition of row representation we have $v^{[2,j]}_{a_2} = 1$. We also know that $F^{[2]}(a_1, *)$ is a nonzero vector and is linearly dependent with $\mathbf{v}^{[2,j]}$. Therefore, we have

$$F^{[1]}(a_1) = \sum_{b \in D} F^{[2]}(a_1, b) = F^{[2]}(a_1, a_2) \cdot \sum_{b \in D} v^{[2,j]}_b.$$

It reduces the computation of $F^{[1]}(a_1)$ to that of $F^{[2]}(a_1, a_2)$. If $n = 2$, then we are already done because $F^{[2]}(a_1, a_2)$ can be evaluated efficiently using the input instance $I$. Otherwise we continue and reduce the computation of $F^{[2]}(a_1, a_2)$ to that of $F^{[3]}(a_1, a_2, a_3)$, for some appropriate $a_3 \in D$.

Because $F^{[2]}(a_1, *)$ is nonzero and is linearly dependent with $\mathbf{v}^{[2,j]}$, we have $F^{[2]}(a_1, a_2) \neq 0$ and thus $(a_1, a_2) \in S^{[3,j]}$ for some $j \in [s_3]$. By using the witness functions $\omega^{[3,j]}$, we can find this $j \in [s_3]$ efficiently and by definition, $F^{[3]}(a_1, a_2, *)$ is nonzero and linearly dependent with $\mathbf{v}^{[3,j]}$. Let $a_3$ denote the smallest nonzero index of $\mathbf{v}^{[3,j]}$, then we have

$$v^{[3,j]}_{a_3} = 1 \quad \text{and} \quad F^{[2]}(a_1, a_2) = \sum_{b \in D} F^{[3]}(a_1, a_2, b) = F^{[3]}(a_1, a_2, a_3) \cdot \sum_{b \in D} v^{[3,j]}_b.$$

This further reduces the computation of $F^{[1]}(a_1)$ to that of $F^{[3]}(a_1, a_2, a_3)$. After a total $n-1$ rounds of such reductions, it suffices to compute $F^{[n]}(a_1, \ldots, a_n)$ for some appropriate $a_2, a_3, \ldots, a_n \in D$, in order to get $F^{[1]}(a_1)$. This gives an efficient algorithm for $F^{[1]}(a_1)$ as $F^{[n]}$ can be evaluated efficiently using $I$.

A formal recursive procedure called ComputeF is described in Figure 1. It takes two inputs: $t$ and $\mathbf{a}$, where $t : 1 \leq t \leq n$ and $\mathbf{a} \in D^t$, and outputs $F^{[t]}(\mathbf{a})$. Its correctness can be easily proved by induction on $t$, and its running time is polynomial because the total number of recursive calls is at most $n - 1$ and in each call, the only non-trivial part is line 4 which has an efficient implementation by Lemma 11. $\square$

## 4 The Block Orthogonality Condition

We prove Lemma 17 in this section. We start with the following hardness lemma for *pure* functions and then use the Purification Lemma to generalize it to general functions:

**Lemma 23.** *Let $F : D^n \to \mathbb{C}$ be a pure function with $n \geq 2$. If $F$ is not block-orthogonal, then* #CSP($F$) *is* #P-*hard.*

*Proof.* Because $F$ is pure, we let $K$ denote the constant order($F$). Without loss of generality, we assume that $F$ is block-rank-1 since otherwise, #CSP($F$) is #P-hard by Lemma 4.

Now assume $F$ is not block-orthogonal. Then by definition, there exist $\mathbf{x}, \mathbf{y} \in D^{n-1}$ such that the two vectors $F(\mathbf{x}, *)$ and $F(\mathbf{y}, *)$ have at least one common non-zero entry, but are neither linearly dependent

ComputeF $(t, \mathbf{a})$, where $t : 1 \le t \le n$ and $\mathbf{a} \in D^t$

1.    if $t = n$ then
2.       use the input instance $I$ to evaluate $F^{[n]}(\mathbf{a}) = F(\mathbf{a})$; output $F(\mathbf{a})$ and exit
3.    end if
4.    use $\omega^{[t+1,j]}$, $j \in [s_{t+1}]$, to check if there is a $j \in [s_{t+1}]$ such that $\mathbf{a} \in S^{[t+1,j]}$
5.    if no such $j \in [s_{t+1}]$ exists then
6.       output $0$ and exit
7.    else
8.       let $j \in [s_{t+1}]$ be the unique index such that $\mathbf{a} \in S^{[t+1,j]}$
9.       let $a_{t+1} \in D$ be the smallest nonzero index of $\mathbf{v}^{[t+1,j]}$
10.      compute $F^{[t+1]}(\mathbf{a}, a_{t+1}) =$ ComputeF $(t+1, \mathbf{a} \circ a_{t+1})$ with a recursive call
11.      output the following and exit

$$F^{[t]}(\mathbf{a}) = \sum_{b \in D} F^{[t+1]}(\mathbf{a}, b) = F^{[t+1]}(\mathbf{a}, a_{t+1}) \cdot \sum_{b \in D} v_b^{[t+1,j]}$$

12.   end if

Figure 1: The recursive procedure ComputeF

nor block-orthogonal. Because $F$ is block-rank-1, we have $F(\mathbf{x}, i) = 0$ if and only if $F(\mathbf{y}, i) = 0$. We use $T \subseteq D$ to denote the nonempty set of $i \in D$ such that $F(\mathbf{x}, i)$ is non-zero. As $F$ is pure and block-rank-1 we can partition $T$ into $T_1, \ldots, T_\ell$ for some $\ell \ge 1$ and there are positive $a, b, \mu_1 > \cdots > \mu_\ell > 0$ such that

$$F(\mathbf{x}, i) = a \cdot \mu_j \cdot c(\mathbf{x}, i) \quad \text{and} \quad F(\mathbf{y}, i) = b \cdot \mu_j \cdot c(\mathbf{y}, i), \quad \text{for all } j \in [\ell] \text{ and } i \in T_j,$$

where $c(\mathbf{x}, i)$ and $c(\mathbf{y}, i)$ are all *roots of unity* whose orders divide $K$.

To show #CSP$(F)$ is #P-hard, we let $\mathbf{A}_r$, for each $r \ge 1$, denote the following $d^{n-1} \times d^{n-1}$ matrix:

$$A_r(\mathbf{w}, \mathbf{w}') = \sum_{i \in D} F(\mathbf{w}, i) \cdot \left(F(\mathbf{w}', i)\right)^{rK-1}, \quad \text{for all } \mathbf{w}, \mathbf{w}' \in D^{n-1},$$

and prove that for every $r \ge 1$ (note that the square matrix $\mathbf{A}_r$ here is not necessarily symmetric),

$$\text{EVAL}(\mathbf{A}_r) \le_T \text{\#CSP}(F).$$

Given any input directed graph $G = (V, E)$ of EVAL$(\mathbf{A}_r)$, we construct $I$ with the following variables:

$$z_{v,1}, \ldots, z_{v,n-1}, w_e, \quad \text{for all } v \in V \text{ and } e \in E,$$

ranging over $D$. Then for each edge $e = uv \in E$, we apply

$$F \text{ over } (z_{u,1}, \ldots, z_{u,n-1}, w_e) \text{ and } (rK - 1) \text{ copies of } F \text{ over } (z_{v,1}, \ldots, z_{v,n-1}, w_e).$$

The reduction then follows because $Z_{\mathbf{A}_r}(G) = Z(F_I)$, where $F_I$ denotes the function that $I$ defines. Now to prove that #CSP($F$) is #P-hard, it suffices to show that EVAL($\mathbf{A}_r$) is #P-hard for some $r \geq 1$.

Focusing on the $2 \times 2$ sub-matrix of $\mathbf{A}_r$ indexed by $\mathbf{x}$ and $\mathbf{y}$, we have

$$A_r(\mathbf{x}, \mathbf{x}) = \sum_{j \in [\ell]} \sum_{i \in T_j} \left(a \cdot \mu_j \cdot c(\mathbf{x}, i)\right)^{rK} = a^{rK} \sum_{j \in [\ell]} |T_j| \cdot (\mu_j)^{rK}$$

and

$$A_r(\mathbf{y}, \mathbf{y}) = \sum_{j \in [\ell]} \sum_{i \in T_j} \left(b \cdot \mu_j \cdot c(\mathbf{y}, i)\right)^{rK} = b^{rK} \sum_{j \in [\ell]} |T_j| \cdot (\mu_j)^{rK}$$

while

$$A_r(\mathbf{x}, \mathbf{y}) = \sum_{j \in [\ell]} \sum_{i \in T_j} a \cdot \mu_j \cdot c(\mathbf{x}, i) \cdot \left(b \cdot \mu_j \cdot c(\mathbf{y}, i)\right)^{rK-1} = a \cdot b^{rK-1} \sum_{j \in [\ell]} (\mu_j)^{rK} \sum_{i \in T_j} c(\mathbf{x}, i) \cdot \overline{c(\mathbf{y}, i)}$$

and

$$A_r(\mathbf{y}, \mathbf{x}) = \sum_{j \in [\ell]} \sum_{i \in T_j} b \cdot \mu_j \cdot c(\mathbf{y}, i) \cdot \left(a \cdot \mu_j \cdot c(\mathbf{x}, i)\right)^{rK-1} = a^{rK-1} \cdot b \sum_{j \in [\ell]} (\mu_j)^{rK} \sum_{i \in T_j} c(\mathbf{y}, i) \cdot \overline{c(\mathbf{x}, i)}$$

We use $L$ to denote

$$\sum_{j \in [\ell]} (\mu_j)^{rK} \sum_{i \in T_j} c(\mathbf{x}, i) \cdot \overline{c(\mathbf{y}, i)}$$

Since all the $\mu_j$'s are positive, we have

$$A_r(\mathbf{x}, \mathbf{y}) \cdot A_r(\mathbf{y}, \mathbf{x}) = a^{rK} \cdot b^{rK} \cdot |L|^2.$$

We discuss the following three cases. First, if

$$|L| = \sum_{j \in [\ell]} |T_j| \cdot (\mu_j)^{rK}, \quad \text{for some } r \geq 1.$$

then by Cauchy-Schwarz, it must be the case that $c(\mathbf{x}, *)$ and $c(\mathbf{y}, *)$, as two $|T|$-dimensional vectors, are linearly dependent and thus, $F(\mathbf{x}, *)$ and $F(\mathbf{y}, *)$ are linearly dependent, contradicting the assumption.

Second, if $L = 0$ for all $r \geq 1$, then by solving a Vandermonde system, it must be the case that

$$\sum_{i \in T_j} c(\mathbf{x}, i) \cdot \overline{c(\mathbf{y}, i)} = 0, \quad \text{for all } j \in [\ell].$$

As a result, these two rows are actually block-orthogonal, contradicting the assumption again.

Otherwise, we must have

$$0 < |L| < \sum_{j \in [\ell]} |T_j| \cdot (\mu_j)^{rK}, \quad \text{for some } r \geq 1.$$

So all the four entries of this sub-matrix of $|\mathbf{A}_r|$ are positive but its rank is 2. This implies that $|\mathbf{A}_r|$ is not a block-rank-1 matrix. By Lemma 3, we have $\mathsf{EVAL}(\mathbf{A}_r)$ is #P-hard and so is $\#\mathsf{CSP}(F)$.

This finishes the proof of the lemma. □

Lemma 17 then follows from Lemma 23, Lemma 16, and Lemma 7.

*Proof of Lemma 17.* Assume $\mathcal{F}$ does not satisfy the Block Orthogonality condition.

Let $\{F_1, \ldots, F_k\} \subset \mathcal{W}_\mathcal{F}$ be a finite set that violates the Block Orthogonality condition. Let

$$(F'_1, \ldots, F'_k) = \mathsf{Pure}(F_1, \ldots, F_k),$$

then by Lemma 7 and Lemma 16, we have for each $i \in [k]$,

$$\#\mathsf{CSP}(F'_i) \leq_T \#\mathsf{CSP}(F'_1, \ldots, F'_k) \equiv_T \#\mathsf{CSP}(F_1, \ldots, F_k) \leq_T \#\mathsf{CSP}(\mathcal{F}).$$

If $F'_i$ is not block-orthogonal, then by Lemma 23, $\#\mathsf{CSP}(F'_i)$ is #P-hard and so is $\#\mathsf{CSP}(\mathcal{F})$. □

## 5 The Type Partition Condition

We prove Lemma 18 in this section.

Again, we start by working on pure functions. Let $F: D^n \to \mathbb{C}$ be a pure function where $n \geq 2$. Also assume that $F$ is block-orthogonal (and in particular, block-rank-1 as well).

Let $\mathcal{S} = \{(S_1, \mathbf{v}_1), \ldots, (S_k, \mathbf{v}_k)\}$ be the row representation of $F$, where $S_1, \ldots, S_k \subseteq D^{n-1}$, and let

$$\Psi = S_1 \cup \cdots \cup S_k.$$

From the pair $(\Psi, (S_1, \ldots, S_k))$, we define the type map $\mathsf{type}(\cdot)$: For any $\ell \in [n-1]$ and any $\mathbf{x} \in D^\ell$,

$$\mathsf{type}(\mathbf{x}) = \Big\{j \in [k] : \exists \mathbf{y} \in S_j \text{ such that } \mathbf{x} = \mathsf{Pr}_{[\ell]} \mathbf{y}\Big\}.$$

We show that if $\mathsf{type}(\cdot)$ is not a type-partition map, then $\#\mathsf{CSP}(F)$ is #P-hard.

**Lemma 24.** *Let $F: D^n \to \mathbb{C}$ be a pure and block-orthogonal function with arity $n \geq 2$, then the problem $\#\mathsf{CSP}(F)$ is #P-hard if there exist an $\ell \in [n-1]$ and $\mathbf{x}, \mathbf{y} \in D^\ell$ such that*

$$\text{neither } \mathsf{type}(\mathbf{x}) \cap \mathsf{type}(\mathbf{y}) = \emptyset \text{ nor } \mathsf{type}(\mathbf{x}) = \mathsf{type}(\mathbf{y}). \tag{13}$$

*Proof.* We start with some notation. Let $K$ be the constant $\mathsf{order}(F)$.

Because $\mathcal{S}$ is the row representation of $F$, there is a function $g: \Psi \to \mathbb{C}$ such that

$$F(\mathbf{x}, *) = g(\mathbf{x}) \cdot \mathbf{v}_j, \quad \text{for all } j \in [k] \text{ and } \mathbf{x} \in S_j.$$

By the definition of row representation, $g(\mathbf{x})$ is exactly the first non-zero entry of $F(\mathbf{x}, *)$.

As $F$ is pure, $g(\mathbf{x})$ is the product of a positive integer and a root of unity whose order divides $K$, for all $\mathbf{x} \in \Psi$. This then implies that all the nonzero entries of $\mathbf{v}_1, \ldots, \mathbf{v}_k$ are products of a positive rational number and a root of unity whose order divides $K$.

Moreover, we know that for any $i \neq j \in [k]$, it follows from Lemma 5 that

$$\sum_{a \in D} v_{i,a} \cdot (v_{j,a})^{K-1} = 0, \tag{14}$$

because they are not only orthogonal but also block-orthogonal. For each $j \in [k]$, we let $c_j > 0$ denote

$$c_j = \sum_{a \in D} v_{j,a} \cdot (v_{j,a})^{K-1} = \sum_{a \in D} |v_{j,a}|^K.$$

Now we start the proof. Let $\ell \in [n-1]$ and let $\mathbf{x}, \mathbf{y} \in D^\ell$ be two vectors that satisfy (13). Note that if $\ell = n-1$, then $\mathsf{type}(\mathbf{x})$ is either the empty set or a singleton set. Therefore, to satisfy (13), $\ell$ has to be smaller than $n - 1$. Without loss of generality, we let

$$\mathsf{type}(\mathbf{x}) = L_1 \cup L_2 \quad \text{and} \quad \mathsf{type}(\mathbf{y}) = L_1 \cup L_3,$$

where $L_1$ and at least one of $L_2, L_3$ are nonempty and they are pairwise disjoint subsets of $[k]$.

Let $\mathbf{A}$ denote the following $d^\ell \times d^\ell$ matrix: For $\mathbf{z}, \mathbf{w} \in D^\ell$, the $(\mathbf{z}, \mathbf{w})^{\text{th}}$ entry of $\mathbf{A}$ is

$$A(\mathbf{z}, \mathbf{w}) = \sum_{\mathbf{z}', \mathbf{w}' \in D^{n-1-\ell}} \left( \sum_{p \in D} F(\mathbf{z}, \mathbf{z}', p) \cdot \left(F(\mathbf{w}, \mathbf{w}', p)\right)^{K-1} \right) \left( \sum_{q \in D} \left(F(\mathbf{z}, \mathbf{z}', q)\right)^{K-1} \cdot F(\mathbf{w}, \mathbf{w}', q) \right)$$

It is easy to see that $\mathbf{A}$ is symmetric. We can then use the following construction to show that

$$\mathsf{EVAL}(\mathbf{A}) \leq_T \#\mathsf{CSP}(F). \tag{15}$$

Given any undirected graph $G = (V, E)$, we construct an instance $I$ with the following variables:

$$v_1, \ldots, v_\ell \text{ for each } v \in |V| \quad \text{and} \quad p_e, q_e, s_{e,\ell+1}, \ldots, s_{e,n-1}, r_{e,\ell+1}, \ldots, r_{e,n-1} \text{ for each } e \in |E|.$$

For each $e = uv \in E$, we apply one copy of $F$ over

$$(u_1, \ldots, u_\ell, s_{e,\ell+1}, \ldots, s_{e,n-1}, p_e) \quad \text{and} \quad (v_1, \ldots, v_\ell, r_{e,\ell+1}, \ldots, r_{e,n-1}, q_e);$$

and apply $(K-1)$ copies of $F$ over

$$(u_1, \ldots, u_\ell, s_{e,\ell+1}, \ldots, s_{e,n-1}, q_e) \quad \text{and} \quad (v_1, \ldots, v_\ell, r_{e,\ell+1}, \ldots, r_{e,n-1}, p_e).$$

It then follows from the construction of $\mathbf{A}$ from $F$ that $Z_\mathbf{A}(G) = Z(F_I)$ where $F_I$ is the function that $I$ defines, and (15) follows. To finish the proof, it now suffices to show that $\mathsf{EVAL}(\mathbf{A})$ is #P-hard.

For this purpose, we analyze the four entries of $\mathbf{A}$ with $\mathbf{z}, \mathbf{w} \in \{\mathbf{x}, \mathbf{y}\}$.

For each $i \in \mathsf{type}(\mathbf{x}) = L_1 \cup L_2$, we let $U_i$ denote the nonempty set of vectors $\mathbf{x}' \in D^{n-\ell-1}$ such that $\mathbf{x} \circ \mathbf{x}' \in S_i$. And we define $V_i$ similarly for $\mathbf{y}$. Then for $i \neq j \in L_1 \cup L_2$ and $\mathbf{z}' \in U_i, \mathbf{w}' \in U_j$, we have

$$\sum_{p \in D} F(\mathbf{x}, \mathbf{z}', p) \cdot \left(F(\mathbf{x}, \mathbf{w}', p)\right)^{K-1} = 0$$

by (14). This can be used to simplify the sum in $A(\mathbf{x}, \mathbf{x})$ as follows:

$$A(\mathbf{x}, \mathbf{x}) = \sum_{i \in L_1 \cup L_2} \sum_{\mathbf{x}', \mathbf{x}'' \in U_i} \left(\sum_{p \in D} g(\mathbf{x}, \mathbf{x}') v_{i,p} \cdot \left(g(\mathbf{x}, \mathbf{x}'') v_{i,p}\right)^{K-1}\right) \left(\sum_{q \in D} \left(g(\mathbf{x}, \mathbf{x}') v_{i,q}\right)^{K-1} \cdot g(\mathbf{x}, \mathbf{x}'') v_{i,q}\right)$$

$$= \sum_{i \in L_1 \cup L_2} \sum_{\mathbf{x}', \mathbf{x}'' \in U_i} \left|g(\mathbf{x}, \mathbf{x}')\right|^K \cdot \left|g(\mathbf{x}, \mathbf{x}'')\right|^K \cdot (c_i)^2 = \sum_{i \in L_1 \cup L_2} \left(\sum_{\mathbf{x}' \in U_i} \left|g(\mathbf{x}, \mathbf{x}')\right|^K \cdot c_i\right)^2$$

Similarly, using the same argument, we have

$$A(\mathbf{y}, \mathbf{y}) = \sum_{i \in L_1 \cup L_3} \sum_{\mathbf{y}', \mathbf{y}'' \in V_i} \left|g(\mathbf{y}, \mathbf{y}')\right|^K \cdot \left|g(\mathbf{y}, \mathbf{y}'')\right|^K \cdot (c_i)^2 = \sum_{i \in L_1 \cup L_3} \left(\sum_{\mathbf{y}' \in V_i} \left|g(\mathbf{y}, \mathbf{y}')\right|^K \cdot c_i\right)^2$$

On the other hand, by a similar proof, we also have

$$A(\mathbf{x}, \mathbf{y}) = \sum_{i \in L_1} \sum_{\mathbf{x}' \in U_i, \mathbf{y}' \in V_i} \left(\sum_{p \in D} g(\mathbf{x}, \mathbf{x}') \cdot v_{i,p} \cdot \left(g(\mathbf{y}, \mathbf{y}') \cdot v_{i,p}\right)^{K-1}\right) \left(\sum_{q \in D} \left(g(\mathbf{x}, \mathbf{x}') \cdot v_{i,q}\right)^{K-1} \cdot g(\mathbf{y}, \mathbf{y}') \cdot v_{i,q}\right)$$

$$= \sum_{i \in L_1} \sum_{\mathbf{x}' \in U_i, \mathbf{y}' \in V_i} \left|g(\mathbf{x}, \mathbf{x}')\right|^K \cdot \left|g(\mathbf{y}, \mathbf{y}')\right|^K \cdot (c_i)^2$$

$$= \sum_{i \in L_1} \left(\sum_{\mathbf{x}' \in U_i} \left|g(\mathbf{x}, \mathbf{x}')\right|^K \cdot c_i\right) \left(\sum_{\mathbf{y}' \in V_i} \left|g(\mathbf{y}, \mathbf{y}')\right|^K \cdot c_i\right)$$

and $A(\mathbf{y}, \mathbf{x}) = A(\mathbf{x}, \mathbf{y})$ (since the construction is symmetric). Because $L_1$ is nonempty, we have

$$A(\mathbf{x}, \mathbf{y}) = A(\mathbf{y}, \mathbf{x}) > 0.$$

It is now easy to see that if at least one of the $L_2, L_3$ is nonempty, then we have

$$A(\mathbf{x}, \mathbf{x}) \cdot A(\mathbf{y}, \mathbf{y}) > A(\mathbf{x}, \mathbf{y}) \cdot A(\mathbf{y}, \mathbf{x}).$$

By Theorem 2, we have that $\mathsf{EVAL}(\mathbf{A})$ is #P-hard and so is $\#\mathsf{CSP}(F)$. This proves the lemma. □

Finally, we use the Purification Lemma to prove Lemma 18.

*Proof of Lemma 18.* Without loss of generality, we may assume that $\mathcal{F}$ satisfies the Block Orthogonality condition (since otherwise, $\#\mathsf{CSP}(\mathcal{F})$ is #P-hard by Lemma 17).

Let $F \in \mathcal{W}_\mathcal{F}$ be a function of arity $\geq 2$, and let $F' = \mathsf{Pure}(F)$. By Lemma 16,

$$\#\mathsf{CSP}(F') \equiv_T \#\mathsf{CSP}(F) \leq_T \#\mathsf{CSP}(\mathcal{F}).$$

As $\mathcal{F}$ satisfies the Block Orthogonality condition, $F'$ is block-orthogonal. From Corollary 1 of the Purification Lemma, $F$ and $F'$ induce the same equivalence relation $\sim_F$ and $\sim_{F'}$ and therefore, the type map $\mathsf{type}_F(\cdot)$ and $\mathsf{type}_{F'}(\cdot)$, induced by $F$ and $F'$, respectively, are the same. If $\mathsf{type}_F(\cdot)$ is not a type-partition map, then neither is $\mathsf{type}_{F'}(\cdot)$. By Lemma 24, $\#\mathsf{CSP}(F')$ is #P-hard and so is $\#\mathsf{CSP}(\mathcal{F})$.

This finishes the proof of the lemma. □

## 6 The Mal'tsev Condition

We prove Lemma 19 in this section. It follows directly from the following lemma:

**Lemma 25.** *If $\mathcal{F}$ satisfies the Block Orthogonality condition, then for any finite $\Gamma \subset \Lambda_\mathcal{F}$, we have*

$$\#\mathsf{CSP}(\Gamma) \leq_T \#\mathsf{CSP}(\mathcal{F}).$$

*Proof of Lemma 19.* If $\mathcal{F}$ does not satisfy the Block Orthogonality condition then we are done by Lemma 17. Assume $\mathcal{F}$ satisfies the Block Orthogonality condition but does not satisfy the Mal'tsev condition.

By Corollary 2, there exists a finite set $\Gamma \subset \Lambda_\mathcal{F}$ with $\#\mathsf{CSP}(\Gamma)$ being #P-hard. Then using Lemma 25, we know $\#\mathsf{CSP}(\mathcal{F})$ is also #P-hard, and the lemma is proven. □

Now we prove Lemma 25.

*Proof of Lemma 25.* Given $\Gamma$, we can find a finite subset $\{F_1, \ldots, F_k\} \subset \mathcal{W}_\mathcal{F}$ such that $\Gamma \subseteq \Delta$, where

$$\Delta = \Big\{ \Phi_i = \mathsf{Boolean}(F_i) : i \in [k] \Big\} \cup \Big\{ \Omega_i : i \in [k] \text{ and the arity of } F_i \text{ is} \geq 2 \Big\}.$$

We use $r_i \geq 1$ to denote the arity of $F_i$. Here $\Omega_i$ is the following relation over $2(r_i - 1)$ variables:

$$(\mathbf{x}, \mathbf{y}) \in \Omega_i \iff F_i(\mathbf{x}, *) \text{ and } F_i(\mathbf{y}, *) \text{ are non-zero and linearly dependent}$$

By Lemma 16, we have
$$\#\mathsf{CSP}(F_1, \ldots, F_k) \leq_T \#\mathsf{CSP}(\mathcal{F}), \tag{16}$$

so we only need to give a polynomial-time reduction from $\#\mathsf{CSP}(\Delta)$ to $\#\mathsf{CSP}(F_1, \ldots, F_k)$.

To this end, we first apply the Purification Lemma to get
$$(F'_1, \ldots, F'_k) = \mathsf{Pure}(F_1, \ldots, F_k). \tag{17}$$

All the new functions $F'_1, \ldots, F'_k$ are pure, and we have
$$\#\mathsf{CSP}(F'_1, \ldots, F'_k) \equiv_T \#\mathsf{CSP}(F_1, \ldots, F_k). \tag{18}$$

We use $K$ to denote the least common multiplier of the orders of all the pure $F'_i$'s.

The plan of the proof is the following. For every $i \in [k]$ with $r_i \geq 2$, we use a construction to define, from $F'_i$, a $2(r_i - 1)$-ary function $H_i$ and prove that

$$\#\mathsf{CSP}\Big(\big\{F'_i : i \in [k]\big\} \cup \big\{H_i : i \in [k] \text{ and } r_i \geq 2\big\}\Big) \leq_\mathsf{T} \#\mathsf{CSP}(F'_1, \ldots, F'_k). \tag{19}$$

We will also show that for every $i \in [k]$ with $r_i \geq 2$,

$$\Omega_i = \mathsf{Boolean}(H_i). \tag{20}$$

On the other hand, by Property 2 of the Purification Lemma, we know that

$$\Phi_i = \mathsf{Boolean}(F_i) = \mathsf{Boolean}(F'_i), \quad \text{for all } i \in [k].$$

As a result, by Lemma 6, we have

$$\#\mathsf{CSP}(\Delta) \leq_\mathsf{T} \#\mathsf{CSP}\Big(\big\{F'_i : i \in [k]\big\} \cup \big\{H_i : i \in [k] \text{ and } r_i \geq 2\big\}\Big), \tag{21}$$

and the lemma follows by combining (21), (19), (18) and (16).

For each $i \in [k]$ with $r_i \geq 2$. We use $H_i$ to denote the following function:

$$H_i(\mathbf{x}, \mathbf{y}) = \sum_{z \in D} F'_i(\mathbf{x}, z) \cdot \big(F'_i(\mathbf{y}, z)\big)^{K-1}, \quad \text{for all } \mathbf{x}, \mathbf{y} \in D^{r_i - 1}.$$

We now use the following construction to show (19). Given any instance $I$ of the first problem in (19) we construct an instance $I'$ as follows. We start with the same set of variables as $I$, and add all the tuples in $I$ whose function is $F'_i$ to $I'$. For each other tuple in $I$, i.e., $(H_i, x_1, \ldots, x_{r_i-1}, y_1, \ldots, y_{r_i-1})$, we create a new variable $z$ and then add the following $K$ tuples to $I'$:

$$\big(F'_i, x_1, \ldots, x_{r_i-1}, z\big) \quad \text{and} \quad (K-1) \text{ copies of } \big(F'_i, y_1, \ldots, y_{r_i-1}, z\big).$$

It is easy to show that $Z(F_I) = Z(F_{I'})$ where $F_I, F_{I'}$ are the functions defined by $I, I'$, and (19) follows.

Finally we prove (20). As $\mathcal{F}$ satisfies the Block Orthogonality condition, (17) implies that $F'_i$ is block-orthogonal. From this, it follows from Lemma 5 that

$$H_i(\mathbf{x}, \mathbf{y}) \neq 0 \iff F'_i(\mathbf{x}, *) \text{ and } F'_i(\mathbf{y}, *) \text{ are nonzero and linearly dependent}$$

Using Corollary 1, the two equivalence relations $\sim_{F_i}$ and $\sim_{F'_i}$, induced by $F_i$ and $F'_i$, respectively, are the same. Therefore, $F'_i(\mathbf{x}, *), F'_i(\mathbf{y}, *)$ are nonzero and linearly dependent if and only if $F_i(\mathbf{x}, *), F_i(\mathbf{y}, *)$ are nonzero and linearly dependent. This proves (20) and finishes the proof of the lemma. $\square$

## 7 Polynomial-Time Operations on Witness Functions

In this section, we present three useful polynomial-time operations on witness functions of relations that share a common Mal'tsev polymorphism. They will then be used later in Section 8 to prove Lemma 21.

## 7.1 Variable Permutation of Witness Functions

**Lemma 26** (Variable Permutation). *Let $\Phi \subseteq D^n$ be an $n$-ary relation. Let $\varphi$ be a Mal'tsev polymorphism of $\Phi$ and $\omega$ be a witness function of $\Phi$. Then given any permutation $\pi$ over $[n]$, we can compute a witness function $\omega'$ for $\pi(\Phi)$ in time polynomial in $n$.*

*Proof.* It suffices to show that, given any $i \in [n-1]$, we can construct a witness function $\omega'$ for

$$\Phi' = \Big\{ (a_1, \ldots, a_i, a_{i+1}, \ldots, a_n) \,\Big|\, (a_1, \ldots, a_{i+1}, a_i, \ldots, a_n) \in \Phi \Big\}.$$

in time polynomial in $n$. For each $j \in [n]$, we use $\sim_j$ and $\sim'_j$ to denote the equivalence relations defined by $\Phi$ and $\Phi'$, respectively. Clearly for $j \neq i$ or $i+1$, $\sim'_j$ is the same as $\sim_j$ and thus, we can set

$$\omega'(j, a) = \omega(j, a), \quad \text{for all } a \in D.$$

Next we compute $\sim'_i$. Let $b \in \Pr_i \Phi' = \Pr_{i+1} \Phi$. Note that the latter can be computed efficiently from $\omega$. We want to compute the class $\mathcal{E}$ of $b$ in $\sim'_i$ and in addition, a witness for each $b' \in \mathcal{E}$ that shares the same $(i-1)$-prefix. We are then done for $\mathcal{E}$ by setting $\omega'(i, b')$ to be this witness for every $b' \in \mathcal{E}$.

To this end, we denote $\omega(i+1, b)$, a witness for $(i+1, b)$ in $\Phi$, by

$$\mathbf{x} \circ a \circ b \circ \mathbf{u} \in \Phi, \quad \text{where } \mathbf{x} \in D^{i-1}, a \in D \text{ and } \mathbf{u} \in D^{n-i-1}. \tag{22}$$

We then use Lemma 13 to compute a witness function for $\Phi(\mathbf{x}, *)$ on $n - (i-1)$ variables, and use it to project $\Phi(\mathbf{x}, *)$ on its second coordinate: $\Pr_2 \Phi(\mathbf{x}, *)$. We now show that $\mathcal{E} = \Pr_2 \Phi(\mathbf{x}, *)$.

Clearly every $b' \in \Pr_2 \Phi(\mathbf{x}, *)$ satisfies $b' \sim'_i b$, because $b' \in \Pr_2 \Phi(\mathbf{x}, *)$ implies that there is a witness for $(i+1, b')$ in $\Phi$ with the same prefix $\mathbf{x}$. Now suppose $b' \sim'_i b$, then by the definition of $\sim'_i$, there exist a $\mathbf{y} \in D^{i-1}$ and $a_1, a_2 \in D$, $\mathbf{u}_1, \mathbf{u}_2 \in D^{n-i-1}$ such that

$$\mathbf{y} \circ a_1 \circ b' \circ \mathbf{u}_1 \in \Phi \quad \text{and} \quad \mathbf{y} \circ a_2 \circ b \circ \mathbf{u}_2 \in \Phi.$$

Applying the Mal'tsev polymorphism $\varphi$ on these two vectors together with the one in (22) then gives a witness for $(i+1, b')$ in $\Phi$ with $\mathbf{x}$ as its prefix. This means that $b' \in \Pr_2 \Phi(\mathbf{x}, *)$.

Now we have computed the equivalence class $\mathcal{E}$ of $b$. We can also use the witness function of $\Phi(\mathbf{x}, *)$ to get a witness for $(i+1, b')$ in $\Phi$ with $\mathbf{x}$ being its prefix. This finishes the construction of $\omega'(i, *)$.

Finally, we work on $\sim'_{i+1}$. Let $a \in \Pr_{i+1} \Phi' = \Pr_i \Phi$, then we want to compute the equivalence class $\mathcal{E}$ of $a$ in $\sim'_{i+1}$. We denote the vector $\omega(i, a)$ by

$$\mathbf{x} \circ a \circ b \circ \mathbf{u} \in \Phi, \quad \text{where } \mathbf{x} \in D^{i-1}, b \in D \text{ and } \mathbf{u} \in D^{n-i-1}. \tag{23}$$

We use Lemma 13 to compute a witness function of $\Phi(\mathbf{x}, *)$ and then, $\Pr_{[2]} \Phi(\mathbf{x}, *)$. For every pair in $(a', b') \in \Pr_{[2]} \Phi(\mathbf{x}, *)$, we also compute a vector in $\Phi(\mathbf{x}, *)$ which starts with $a'$ and $b'$. We then collect all the $a' \in D$ such that for some $b' \in D$, both $(a', b'), (a, b') \in \Pr_{[2]} \Phi(\mathbf{x}, *)$, and claim that this is exactly $\mathcal{E}$.

First, it is easy to see that if $(a', b'), (a, b') \in \Pr_{[2]} \Phi(\mathbf{x}, *)$ for some $b' \in D$, then $a \sim'_{i+1} a'$. Conversely, if $a \sim'_{i+1} a'$, then there are $\mathbf{y} \in D^{i-1}$ and $c \in D, \mathbf{u}_1, \mathbf{u}_2 \in D^{n-i-1}$ such that

$$\mathbf{y} \circ a \circ c \circ \mathbf{u}_1 \in \Phi \quad \text{and} \quad \mathbf{y} \circ a' \circ c \circ \mathbf{u}_2 \in \Phi.$$

Applying the Mal'tsev polymorphism $\varphi$ on these two vectors together with the one in (23) then gives a vector in $\Phi$ with prefix $\mathbf{x} \circ a' \circ b$. This implies that $(a', b) \in \mathsf{Pr}_{[2]} \Phi(\mathbf{x}, *)$.

We have computed the class $\mathcal{E}$ of $a$ in $\sim'_{i+1}$. For each $a' \in \mathcal{E}$ with $(a', b'), (a, b') \in \mathsf{Pr}_{[2]} \Phi(\mathbf{x}, *)$, we can compute two vectors in $\Phi$ with prefixes $\mathbf{x} \circ a' \circ b'$ and $\mathbf{x} \circ a \circ b'$. Applying the Mal'tsev polymorphism $\varphi$ on these two vectors together with the one in (23) gives a vector in $\Phi$ with prefix $\mathbf{x} \circ a' \circ b$. As a result, we obtain a witness of $(i+1, a')$ in $\Phi'$, for every $a' \in \mathcal{E}$, which shares the same prefix $\mathbf{x} \circ b$. We then set $\omega'(i+1, a')$ to be this witness for each $a' \in \mathcal{E}$, and this finishes the construction of $\omega'(i+1, *)$. □

## 7.2 Union of Witness Functions

Let $\Psi_1, \ldots, \Psi_s$ be $s$ pairwise disjoint relations over $n$ variables $x_1, \ldots, x_n \in D$. Also assume that $\varphi$ is a Mal'tsev polymorphism of all the $\Psi_k$'s. Let

$$\Phi = \Psi_1 \cup \cdots \cup \Psi_s.$$

In general, $\varphi$ may not be a Mal'tsev polymorphism of $\Phi$. However, the following lemma shows that if it is assumed that $\varphi$ is a Mal'tsev polymorphism of $\Phi$ as well, then we can construct a witness function of $\Phi$ from witness functions of the $\Psi_k$'s efficiently.

**Lemma 27.** *Let $\Psi_1, \ldots, \Psi_s$ be pairwise disjoint and nonempty subsets of $D^n$, and let $\Phi = \Psi_1 \cup \cdots \cup \Psi_s$. Also assume that $\varphi$ is a Mal'tsev polymorphism of both $\Phi$ and the $\Psi_k$'s. Given a witness function $\omega_k$ of $\Psi_k$ for each $k \in [s]$, we can construct a witness function $\omega$ of $\Phi$ in polynomial time (in $s$ and $n$).*

*Proof.* Pick any pair $(i, a) \in [n] \times D$. We first decide whether there is a vector $\mathbf{x} \in \Phi$ such that $x_i = a$. As $\Phi$ is the union of the $\Psi_k$'s, it suffices to check if $\omega_k(i, a) \neq \bot$ for some $k \in [s]$. If $\omega_k(i, a) = \bot$ for every $k \in [s]$, then we simply set $\omega(i, a) = \bot$; otherwise we have computed a witness in $\Phi$ for $(i, a)$.

Next for each $i \in [n]$ we compute the equivalence relation $\sim_i$ of $\Phi$ as follows. Pick any $a \neq b \in D$ for which we have already found witnesses $\mathbf{x}, \mathbf{y}$ in $\Phi$, with $x_i = a$ and $y_i = b$. By Lemma 10, we have

$$a \sim_i b \iff \exists \mathbf{z} \in \Phi \text{ such that } \mathsf{Pr}_{[i]} \mathbf{z} = (x_1, \ldots, x_{i-1}, b)$$

Because $\Phi$ is the union of the $\Psi_k$'s, it happens if and only if there exists such a $\mathbf{z} \in \Psi_k$ for some $k \in [s]$. To check whether $\Psi_k$ has such a $\mathbf{z}$, by Lemma 13, we can use $\omega_k$ to construct a witness function for

$$\Psi_k(x_1, \ldots, x_{i-1}, b, *).$$

Then $\Psi_k$ has a $\mathbf{z}$ with $\mathsf{Pr}_{[i]} \mathbf{z} = (x_1, \ldots, x_{i-1}, b)$ if and only if the witness function we get is nonempty.

It is clear that the computation of $\sim_i$, $i \in [n]$, also gives us a witness function $\omega$ for $\Phi$. □

## 7.3 Splitting a Witness Function, with the Type Partition Condition

Here we describe the *inverse* of the union operation described above. The setting is the following.

Let $\Phi \subseteq D^n$ be a nonempty relation over $n$ variables. And let $\Psi_1, \ldots, \Psi_s$ be an $s$-way partition of $\Phi$, for some $s \in [d]$: The $\Psi_i$'s are nonempty, pairwise disjoint, and satisfy

$$\Phi = \Psi_1 \cup \cdots \cup \Psi_s.$$

Assume that $\varphi$ is a Mal'tsev polymorphism of $\Phi$ and all the $\Psi_i$'s.

At the beginning, we have completely no information about the sets $\Psi_i$'s. Even the number $s$ of sets is not given, though we do know that $s \in [d]$. The only resources we have are a witness function $\omega$ for $\Phi$ and a black box to query: We can send any $\mathbf{x} \in \Phi$ to the black box and it returns the unique $k \in [s]$ such that $\mathbf{x} \in \Psi_k$. The question then is: Can we use $\omega$ and the black box to compute $s \in [d]$ as well as a witness function $\omega_k$ for each $\Psi_k$ in polynomial time and only using polynomially many queries?

In general, we do not know how to solve this problem efficiently. But if the following condition holds then it has an efficient algorithm. Here given any permutation $\pi$ on $[n]$, we use $\mathsf{type}_\pi$ to denote the type map defined by $(\pi(\Phi), (\pi(\Psi_1), \ldots, \pi(\Psi_s)))$, that is,

$$\mathsf{type}_\pi(\mathbf{x}) = \left\{ k \in [s] : \exists\, \mathbf{y} \in \pi(\Psi_k) \text{ such that } \mathsf{Pr}_{[\ell]}\mathbf{y} = \mathbf{x} \right\}, \quad \text{for all } \mathbf{x} \in D^\ell \text{ with } \ell \in [n].$$

We also set $\mathsf{type}_\pi(\epsilon) = [s]$, where $\epsilon$ denotes the empty string.

Our condition requires that $\mathsf{type}_\pi(\cdot)$ is a type-partition map for all permutations $\pi$ over $[n]$:

**Lemma 28.** *Let $\Psi_1, \ldots, \Psi_s$ be an $s$-way partition of $\Phi \subseteq D^n$, for some $s \in [d]$. Assume $\varphi$ is a Mal'tsev polymorphism of $\Phi$ and the $\Psi_k$'s, and $\mathsf{type}_\pi(\cdot)$ defined above is a type-partition map for any permutation $\pi$ on $[n]$. Then given a witness function $\omega$ of $\Phi$ and a black box specified above, we can compute $s \in [d]$ as well as a witness function $\omega_k$ of each $\Psi_k$ in polynomial time and using polynomially many queries (in $n$).*

We start with some definitions and lemmas.

We use $\mathsf{type}(\cdot)$ to denote $\mathsf{type}_\pi(\cdot)$ with $\pi$ being the identity permutation for short, and use

$$\mathfrak{T} = (\mathcal{T}_0, \mathcal{T}_1, \ldots, \mathcal{T}_n), \quad \text{where } \mathcal{T}_j = \left\{ \mathsf{type}(\mathbf{x}) \subseteq [s] : \mathbf{x} \in \mathsf{Pr}_{[j]}\Phi \right\}$$

to denote the list of types of $\mathsf{type}(\cdot)$. Since $\mathsf{type}(\cdot)$ is a type-partition map, we have $|\mathcal{T}_j| \leq s \leq d$ for all $j$. It is clear that all the $\mathcal{T}_j$'s are nonempty because $\Phi$ is nonempty; every set in $\mathcal{T}_j$ is a nonempty subset of $[s]$ because we are only interested in $\mathbf{x} \in \mathsf{Pr}_{[j]}\Phi$ in the definition above.

We need the following definition in the algorithm:

**Definition 8.** *We say $\mathfrak{S} = (\mathcal{S}_0, \mathcal{S}_1, \ldots, \mathcal{S}_n)$ is a* partial list *of $\mathfrak{T}$ if $\mathcal{S}_j \subseteq \mathcal{T}_j$ for all $j : 0 \leq j \leq n$.*

*Given $U \in \mathcal{T}_\ell$ for some $0 \leq \ell \leq n$, we say $\mathfrak{S}$ is* closed with respect to $U$ at level $\ell$*, if $U \in \mathcal{S}_\ell$ and for every $j > \ell$, we have $V \in \mathcal{S}_j$ for any $V \in \mathcal{T}_j$ with $V \subseteq U$.*

*Finally we say $\mathfrak{S}$ is* closed *if it is closed with respect to every $U \in \mathcal{S}_j$ at level $j$, for all $j : 0 \leq j \leq n$.*

In particular, $\mathfrak{S}$ is closed if $\mathcal{S}_j = \emptyset$ for all $j$. It is easy to show from the definition that

**Lemma 29.** *If $\mathfrak{S}$ is a closed partial list of $\mathfrak{T}$ and $\mathcal{S}_0$ has the set $[s]$, then $\mathfrak{S} = \mathfrak{T}$.*

*Proof.* We use induction on $j = 0, 1, \ldots, n$. The base case is trivial as by assumption, $\mathcal{S}_0 = \mathcal{T}_0 = \{[s]\}$.

Now assume that $\mathcal{S}_j = \mathcal{T}_j$ for some $j \geq 0$. As $\mathfrak{S}$ is a partial list of $\mathfrak{T}$, to show $\mathcal{S}_{j+1} = \mathcal{T}_{j+1}$, it suffices to prove that $V \in \mathcal{S}_{j+1}$ for any $V \in \mathcal{T}_{j+1}$. By the earlier discussion on the tree structure of $\mathfrak{T}$ in Section 2.9, there is a unique set $U \in \mathcal{T}_j$ such that $V \subseteq U$. By the inductive hypothesis, $U \in \mathcal{T}_j = \mathcal{S}_j$ and hence, $V \in \mathcal{S}_{j+1}$ as $\mathfrak{S}$ is closed. This finishes the induction and the proof of the lemma. $\square$

We present a recursive procedure $\mathsf{ComputeType}$ for computing both $s \in [d]$ and $\mathfrak{T}$, using the witness function $\omega$ of $\Phi$ and the black box. The procedure is given formally in Figure 2. It takes two inputs:

(i) a vector $\mathbf{x} \in \mathsf{Pr}_{[\ell]}\Phi$, where $\ell : 0 \leq \ell \leq n$ (and $\mathbf{x} = \epsilon$ when $\ell = 0$); and

(ii) a *closed partial list* $\mathfrak{S} = (\mathcal{S}_0, \mathcal{S}_1, \ldots, \mathcal{S}_n)$ of $\mathfrak{T}$. (Note that during the execution of ComputeType, it sometimes updates the list $\mathfrak{S}$ by adding new sets to the $\mathcal{S}_i$'s.)

We analyze the procedure ComputeType and prove the following lemma:

**Lemma 30.** *Let $\mathbf{x} \in \mathsf{Pr}_{[\ell]}\Phi$ for some $\ell : 0 \leq \ell \leq n$ and let $\mathfrak{S}$ be a partial list of $\mathfrak{T}$, then we have*

$$\mathsf{ComputeType}(\mathbf{x}, \mathfrak{S}) = \mathsf{type}(\mathbf{x}).$$

*Let $\mathfrak{S}' = (\mathcal{S}'_0, \mathcal{S}'_1, \ldots, \mathcal{S}'_n)$ denote the tuple $\mathfrak{S}$ after the execution of $\mathsf{ComputeType}(\mathbf{x}, \mathfrak{S})$, then we have*

$$\mathsf{type}(\mathbf{x}) \in \mathcal{S}'_\ell \quad \text{and} \quad \mathcal{S}_j \subseteq \mathcal{S}'_j, \quad \text{for all } j : 0 \leq j \leq n. \tag{24}$$

*Moreover, if $\mathfrak{S}$ is a closed partial list of $\mathfrak{T}$, then so is $\mathfrak{S}'$.*

*Proof.* We prove the lemma by induction on $\ell = n, n-1, \ldots, 1, 0$. The base case when $\ell = n$ is trivial as $\mathfrak{S}$ remains a closed partial list of $\mathfrak{T}$ if a singleton set $\{k\} \in \mathcal{T}_n$ is added to $\mathcal{S}_n$.

Assume the lemma holds for all calls to ComputeType with vectors $\mathbf{x}$ of length $\ell + 1, \ldots, n$, for some $\ell \geq 0$. We now show that if $\mathbf{x} \in \mathsf{Pr}_{[\ell]}\Phi$ and $\mathfrak{S}$ is a partial list of $\mathfrak{T}$, then

$$\mathsf{ComputeType}(\mathbf{x}, \mathfrak{S}) = \mathsf{type}(\mathbf{x})$$

and the new tuple $\mathfrak{S}'$ after the execution satisfies (24). In addition, if $\mathfrak{S}$ is closed then so is $\mathfrak{S}'$.

There are two cases to discuss. First if the algorithm reaches line 9 then we clearly have $\mathsf{type}(\mathbf{x}) = U$ since $\mathsf{type}(\cdot)$ is a type-partition map and the input tuple $\mathfrak{S}$ is assumed to be a partial list of $\mathfrak{T}$. Also the properties about $\mathfrak{S}'$ hold because $\mathfrak{S}' = \mathfrak{S}$ in this case.

Otherwise, the algorithm uses the for-loop to get $U_a$ for each $a \in \mathsf{Pr}_1\Phi'$. By the inductive hypothesis, we know that at the end of each iteration of line 12, $\mathfrak{S}$ remains a partial list of $\mathfrak{T}$. After the for-loop, we have $U_a = \mathsf{type}(\mathbf{x} \circ a)$ and $\mathfrak{S}$ is a partial list with $\mathsf{type}(\mathbf{x} \circ a) \in \mathcal{S}_{\ell+1}$ for all $a \in \mathsf{Pr}_1\Phi'$.

As a result, we have by line 18 and line 21 that

$$\mathsf{ComputeType}(\mathbf{x}, \mathfrak{S}) = U = \bigcup_{a \in \mathsf{Pr}_1\Phi'} U_a = \bigcup_{a \in \mathsf{Pr}_1\Phi'} \mathsf{type}(\mathbf{x} \circ a) = \mathsf{type}(\mathbf{x}).$$

It is easy to show that after the execution, $\mathfrak{S}'$ remains a partial list of $\mathfrak{T}$ and satisfies (24). Next, assume that the input $\mathfrak{S}$ is closed. By the inductive hypothesis, $\mathfrak{S}$ remains closed before line 22 and we have

$$\mathsf{type}(\mathbf{x} \circ a) \in \mathcal{S}_{\ell+1}, \quad \text{for all } a \in \mathsf{Pr}_1\Phi'.$$

Therefore, before and after line 22 $\mathfrak{S}$ is closed with respect to $\mathsf{type}(\mathbf{x} \circ a)$ at level $\ell + 1$ for all $a \in \mathsf{Pr}_1\Phi'$. Also notice that these are all the subsets of $\mathsf{type}(\mathbf{x})$ in $\mathcal{T}_{\ell+1}$. It follows that $\mathfrak{S}$ remains closed after adding $\mathsf{type}(\mathbf{x})$ to $\mathcal{S}_\ell$ in line 22, because $\mathfrak{S}$ remains closed with respect to $\mathsf{type}(\mathbf{x})$ at level $\ell$.

This finishes the induction and the proof of the lemma. □

*Proof of Lemma 28.* We start the proof of Lemma 28 now.

ComputeType$(\mathbf{x}, \mathfrak{S})$, where $\mathbf{x} \in \mathsf{Pr}_{[\ell]}\Phi$ and $\ell : 0 \le \ell \le n$

1.     if $\ell = n$ then
2.         query the black box to get the $k \in [d]$ such that $\mathbf{x} \in \Psi_k$
3.         add $\{k\}$ to $\mathcal{S}_n$ if $\{k\} \notin \mathcal{S}_n$; output $\{k\}$ and exit
4.     end if
5.     compute a witness function $\omega'$ of $\Phi' = \Phi(x_1, \ldots, x_\ell, *)$ ($\Phi' = \Phi$ if $\mathbf{x} = \epsilon$ and $\ell = 0$)
6.     use $\omega'$ to find a vector $\mathbf{y} \in D^{n-\ell}$ such that $\mathbf{x} \circ \mathbf{y} \in \Phi$
7.     query the black box to get the $k \in [d]$ such that $\mathbf{x} \circ \mathbf{y} \in \Psi_k$
8.     if $k$ belongs to one of the subsets $U$ in $\mathcal{S}_\ell$ then
9.         output $U$ and exit
10.    else
11.         use $\omega'$ to compute $\mathsf{Pr}_1\Phi' = \{b \in D : \omega'(1, b) \ne \bot\}$
12.         for each $a \in \mathsf{Pr}_1\Phi'$
13.             let $\mathbf{z} = \omega'(1, a) \in D^{n-\ell}$ (and we have $z_1 = a$ and $\mathbf{x} \circ \mathbf{z} \in \Phi$)
14.             query the black box to get the $k \in [d]$ such that $\mathbf{x} \circ \mathbf{z} \in \Psi_k$
15.             if $k$ belongs to one of the subsets in $\mathcal{S}_{\ell+1}$ then
16.                 denote this subset of $\mathcal{S}_{\ell+1}$ by $U_a$ (and we have $\mathsf{type}(\mathbf{x} \circ a) = U_a$)
17.             else
18.                 let $U_a = $ ComputeType$(\mathbf{x} \circ a, \mathfrak{S})$
19.             end if
20.         end for
21.         let $U = \bigcup_{a \in \mathsf{Pr}_1\Phi'} U_a$
22.         set $\mathcal{S}_\ell$ to be $\mathcal{S}_\ell \cup \{U\}$; output $U$ and exit
23.    end if

Figure 2: The recursive procedure ComputeType

From Lemma 30, we can call ComputeType$(\epsilon, \mathfrak{S})$ with $\mathcal{S}_j = \emptyset$ in $\mathfrak{S}$ for all $j$, to get the number $s \in [d]$ of $\Psi_k$'s as it outputs $\mathsf{type}(\epsilon) = [s]$. By the end of its execution, we also have $\mathsf{type}(\epsilon) \in \mathcal{S}_0$ and $\mathfrak{S}$ remains a closed partial list of $\mathfrak{T}$. It then follows from Lemma 29 that $\mathfrak{S}$ becomes exactly $\mathfrak{T}$.

Next, we show that ComputeType$(\epsilon, \mathfrak{S})$ actually runs in polynomial time, and only uses polynomially

1. use $\omega$ to find a vector $\mathbf{y} \in D^{n-\ell}$ such that $\mathbf{x} \circ \mathbf{y} \in \Phi$
2. query the black box to get the integer $k \in [s]$ such that $\mathbf{x} \circ \mathbf{y} \in \Psi_k$
3. use $k$ to find a subset $U$ in $\mathcal{T}_\ell$ such that $k \in U$

Figure 3: Computation of $\mathsf{type}(\mathbf{x})$ using $\mathfrak{T}$

1. for every $a \in D$ such that $\mathbf{x} \circ a \in \mathsf{Pr}_{[\ell+1]}\Phi$
2.     compute $\mathsf{type}(\mathbf{x}, a)$
3.     if $k \in \mathsf{type}(\mathbf{x}, a)$, then recursively find a $\mathbf{z}$ such that $\mathbf{x} \circ a \circ \mathbf{z} \in \Psi_k$, and output $a \circ \mathbf{z}$
4. end for

Figure 4: Finding a $\mathbf{y}$ such that $\mathbf{x} \circ \mathbf{y} \in \Psi_k$, where $k \in \mathsf{type}(\mathbf{x})$

many queries to the black box. Notice that the running time and number of queries used in each call to ComputeType, excluding those spent in the recursive calls in line 18, are bounded by a polynomial in $n$.

We now show the following statement: during each recursive call to ComputeType in line 18, at least one new set is added to one of the $\mathcal{S}_j$'s in $\mathfrak{S}$. This is because each recursive call to ComputeType in line 18 has the following property: The index $k$ obtained in line 14 belongs to $\mathsf{type}(\mathbf{x} \circ a)$ by the definition of $\mathbf{z}$ in line 13 and the definition of $k$ in line 14. The fact that we reach line 18 means that the condition in line 15 fails and thus, $k \notin$ any set in $\mathcal{S}_{\ell+1}$ before the execution of ComputeType in line 18. But after the execution of ComputeType in line 18, $k \in$ the set $\mathsf{type}(\mathbf{x} \circ a) \in$ the updated $\mathcal{S}_{\ell+1}$. The statement follows.

As a result, each recursive call of ComputeType in line 18 strictly increases the cardinality of $\mathfrak{S}$, but

$$\sum_{\ell=0}^{n} |\mathcal{S}_\ell| \le \sum_{\ell=0}^{n} |\mathcal{T}_\ell| \le 1 + dn = O(n),$$

because $|\mathcal{T}_\ell| \le d$ for every $\ell \in [n]$ and $|\mathcal{T}_0| = 1$. Hence, there can be at most $O(n)$ recursive calls in every execution of $\mathsf{ComputeType}(\mathbf{x}, \mathfrak{S})$. Therefore, we conclude that the total running time and the number of queries to the black box used by $\mathsf{ComputeType}(\epsilon, \mathfrak{S})$ is polynomial in $n$.

We have computed $s \in [d]$ and $\mathfrak{T}$. With $\mathfrak{T}$, we can compute $\mathsf{type}(\mathbf{x})$ for any $\mathbf{x} \in \mathsf{Pr}_{[\ell]}\Phi$ in polynomial time. The algorithm is described in Figure 3. Since $\mathsf{type}(\cdot)$ is a type-partition map, by the definition of $\mathfrak{T}$, we know there is a unique $U \in \mathcal{T}_\ell$ such that $k \in U$, and we have $\mathsf{type}(\mathbf{x}) = U$.

Moreover, given any $\mathbf{x} \in \mathsf{Pr}_{[\ell]}\Phi$ and $k \in \mathsf{type}(\mathbf{x})$, we can find recursively and in polynomial time a $\mathbf{y}$ such that $\mathbf{x} \circ \mathbf{y} \in \Psi_k$. The algorithm is described in Figure 4.

Let $\pi$ be any permutation on $[n]$. By the assumption of the lemma, $\mathsf{type}_\pi(\cdot)$ is also a type-partition map. We note that all the algorithms in Figure 2, 3 and 4 still work correctly, even if we replace $\mathsf{type}(\cdot)$ by $\mathsf{type}_\pi(\cdot)$ and replace the witness function $\omega$ of $\Phi$ by a witness function $\omega_\pi$ of $\pi(\Phi)$. Also note that $\omega_\pi$ can be computed from $\omega$ efficiently using Lemma 26.

Now pick any $k \in [s]$, and we start to construct a witness function $\omega_k$ for $\Psi_k$.

Pick any pair $(i, a)$ with $i \in [n]$ and $a \in D$. We use $\pi$ to denote a permutation over $[n]$ with $\pi(i) = 1$.

Then by using the first two algorithms in Figure 2 and 3, we can compute $\mathsf{type}_\pi(a)$. We use $\mathsf{type}_\pi(a)$ to determine if $a \in \mathsf{Pr}_i \Psi_k$ as follows. If $k \in \mathsf{type}_\pi(a)$, then $a \in \mathsf{Pr}_i \Psi_k$, and we use the algorithm in Figure 4 to find a witness in $\Psi_k$ for $(i, a)$; Otherwise, no such witness exists and we set $\omega_k(i, a) = \bot$.

To derive the equivalence relation $\sim_i$ defined by $\Psi_k$, we pick $a, b \in \mathsf{Pr}_i \Psi_k$ and use $\mathbf{x}, \mathbf{y} \in \Psi_k$ to denote the witnesses in $\Psi_k$ we have found for $(i, a)$ and $(i, b)$. Then we use the algorithm in Figure 3 to check if

$$k \in \mathsf{type}\Big(\big(\mathsf{Pr}_{[i-1]} \mathbf{x}\big) \circ b\Big).$$

It is then easy to show that $a \sim_i b$ if and only if $k \in \mathsf{type}\big((\mathsf{Pr}_{[i-1]}\mathbf{x}) \circ b\big)$. This gives us the relation $\sim_i$.

Finally, we can also use the algorithm in Figure 4 to find a vector $\mathbf{x}'$, for each $b \sim_i a$, such that

$$\big(\mathsf{Pr}_{[i-1]}\mathbf{x}\big) \circ b \circ \mathbf{x}' \in \Psi_k.$$

This finishes the construction of $\omega_k$, and the proof of Lemma 28. $\square$

## 8 Proof of Lemma 21

With all the operations for witness functions developed in the last section, we now prove Lemma 21.

For each $\ell : 2 \le \ell \le n$, we let $\Phi_\ell = \mathsf{Boolean}(F^{[\ell]})$, and use

$$\Big\{\big(S^{[\ell,j]}, \mathbf{v}^{[\ell,j]}\big) : j \in [s_\ell]\Big\}$$

to denote the row representation of $F^{[\ell]}$.

We now show how to compute efficiently $s_\ell \in [d]$, a witness function $\omega_\ell$ for $\Phi_\ell$, and

$$\Big\{\big(\omega^{[\ell,j]}, \mathbf{v}^{[\ell,j]}\big) : j \in [s_\ell]\Big\}$$

in polynomial time for all $2 \le \ell \le n$ such that $\omega^{[\ell,j]}$ is a witness function of $S^{[\ell,j]}$ for all $\ell$ and $j$. Here it makes sense to talk about witness functions for $\Phi_\ell$ and $S^{[\ell,j]}$ since by the Mal'tsev condition and Lemma 20, $\varphi$ is a Mal'tsev polymorphism of all these sets.

We use induction on $\ell = n, \ldots, 2$. We start with the base case $\ell = n$. Let $\mathcal{F} = \{f_1, \ldots, f_h\}$ and let $\varphi$ denote a Mal'tsev polymorphism shared by relations in $\Lambda_\mathcal{F}$ and thus, $\varphi$ is a Mal'tsev polymorphism of

$$\Big\{\mathsf{Boolean}(f_1), \ldots, \mathsf{Boolean}(f_h)\Big\}.$$

Therefore, by Theorem 4, we can construct a witness function $\omega_n$ efficiently for

$$\Phi_n = \mathsf{Boolean}(F^{[n]}) = \mathsf{Boolean}(F),$$

because $\Phi_n = \mathsf{Boolean}(F)$ is the relation defined by an input instance of the unweighted

$$\#\mathsf{CSP}\Big(\mathsf{Boolean}(f_1), \ldots, \mathsf{Boolean}(f_h)\Big).$$

Using $\omega_n$, we can construct a witness function $\omega'_n$ for $\Psi_n = \Pr_{[n-1]} \Phi_n$ using Lemma 14. Because

$$\left\{ \left(S^{[n,j]}, \mathbf{v}^{[n,j]}\right) : j \in [s_n] \right\}$$

denotes the row representation, we have

$$\Psi_n = \bigcup_{j \in [s_n]} S^{[n,j]}.$$

Now let $\pi$ be any permutation from $[n-1]$ to itself. We define the following type map $\mathsf{type}_\pi(\cdot)$:

$$\mathsf{type}_\pi(\mathbf{x}) = \left\{ j \in [s_n] : \exists\, \mathbf{y} \in \pi\bigl(S^{[n,j]}\bigr) \text{ such that } \Pr_{[r]} \mathbf{y} = \mathbf{x} \right\}, \quad \text{for } \mathbf{x} \in D^r \text{ and } r \in [n-1].$$

By the Type Partition condition, we know that $\mathsf{type}_\pi(\cdot)$ is a type-partition map, for any permutation $\pi$. This follows from the fact that, given any function in $\mathcal{W}_\mathcal{F}$, we can arbitrarily permute its variables and the new function still belongs to $\mathcal{W}_\mathcal{F}$.

Therefore, we can now use Lemma 28 to compute $s_n \in [d]$, and construct a witness function $\omega^{[n,j]}$ for each $S^{[n,j]}$, $j \in [s_n]$. Notice that the black box that Lemma 28 needs to query can be implemented quite trivially here: given any $\mathbf{x} \in D^{n-1}$, we can evaluate the vector $F(\mathbf{x}, *)$ efficiently using the input instance $I$. The black box keeps all the linearly independent vectors $F(\mathbf{x}, *)$ evaluated so far and associates each of them with a unique label $j \in [s_n]$. With $\omega^{[n,j]}$ computed, we can use it to get a vector $\mathbf{x} \in S^{[n,j]}$, and then evaluate $F(\mathbf{x}, *)$ to get the representative vector $\mathbf{v}^{[n,j]}$.

Assume for induction that for some $\ell : 2 \le \ell < n$, we have computed $s_t$, a witness function of $\Phi_t$ and

$$\left\{ \left(\omega^{[t,j]}, \mathbf{v}^{[t,j]}\right) : j \in [s_t] \right\}$$

for all $t = \ell + 1, \ldots, n$ such that $\omega^{[t,j]}$ is a witness function for $S^{[t,j]}$. To work on $F^{[\ell]}$, we notice that

$$F^{[\ell]}(\mathbf{x}) = \sum_{a \in D} F^{[\ell+1]}(\mathbf{x}, a).$$

As a result, we have $F^{[\ell]}(\mathbf{x}) \ne 0$ if and only if $\mathbf{x} \in S^{[\ell+1,j]}$ for some $j \in [s_{\ell+1}]$ and

$$\sum_{a \in D} v_a^{[\ell+1,j]} \ne 0.$$

We let $L$ denote the subset of $[s_{\ell+1}]$ such that $j \in L$ if the sum above is nonzero, then we have

$$\Phi_\ell = \mathsf{Boolean}\bigl(F^{[\ell]}\bigr) = \bigcup_{j \in L} S^{[\ell+1,j]}.$$

Using the Mal'tsev condition and Lemma 20 we also know that $\varphi$ is a Mal'tsev polymorphism of $\Phi_\ell$ and the $S^{[\ell+1,j]}$'s. By using Lemma 27 as well as the witness functions $\omega^{[\ell+1,j]}$ for $S^{[\ell+1,j]}$, we can compute a witness function $\omega_\ell$ of $\Phi_\ell$ efficiently.

Next, we use $\omega_\ell$ and Lemma 14 to construct a witness function $\omega'_\ell$ for

$$\Psi_\ell = \Pr_{[\ell-1]} \Phi_\ell,$$

a relation over $\ell - 1$ variables. Because

$$\left\{\left(S^{[\ell,j]}, \mathbf{v}^{[\ell,j]}\right) : j \in [s_\ell]\right\}$$

denotes the row representation of $F^{[\ell]}$, we have

$$\Psi_\ell = \bigcup_{j \in [s_\ell]} S^{[\ell,j]}.$$

Similarly, given any permutation $\pi$ from $[\ell - 1]$ to itself, we define the following type map:

$$\mathsf{type}_\pi(\mathbf{x}) = \left\{j \in [s_\ell] : \exists\, \mathbf{y} \in \pi\bigl(S^{[\ell,j]}\bigr) \text{ such that } \mathsf{Pr}_{[r]}\mathbf{y} = \mathbf{x}\right\}, \quad \text{for } \mathbf{x} \in D^r \text{ and } r \in [\ell - 1].$$

By the Type Partition condition again, $\mathsf{type}_\pi(\cdot)$ is a type-partition map for any permutation $\pi$.

But before we can finally use the algorithm of Lemma 28 to compute $s_\ell \in [d]$ and construct a witness function $\omega^{[\ell,j]}$ for each $S^{[\ell,j]}$ in the row representation, we need to first show how to implement the black box efficiently. For this purpose, it suffices to give an efficient algorithm for computing $F^{[\ell]}(\mathbf{x})$, $\mathbf{x} \in D^\ell$.

This can be done by calling $\mathsf{ComputeF}(\ell, \mathbf{x})$, the polynomial-time algorithm described in the proof of Lemma 22 in Figure 1. Notice that the execution of $\mathsf{ComputeF}(\ell, \mathbf{x})$ uses $s_{\ell+1}, \ldots, s_n$ and the pairs

$$\left\{\left(\omega^{[t,j]}, \mathbf{v}^{[t,j]}\right) : \ell + 1 \leq t \leq n \text{ and } j \in [s_t]\right\},$$

all of which have already been computed by the inductive hypothesis. Now we can use the algorithm in Lemma 28 to compute $s_\ell$ and the pairs $\left(\omega^{[\ell,j]}, \mathbf{v}^{[\ell,j]}\right)$.

This finishes the induction and the lemma is proven.

## 9 Conclusions

We proved a complexity dichotomy theorem for #CSP with complex weights. To this end, we introduced three criteria over the language $\mathcal{F}$: the Block Orthogonality condition, the Type Partition condition and the Mal'tsev condition. We show that #CSP($\mathcal{F}$) is #P-hard if $\mathcal{F}$ violates any of these three conditions, and give a polynomial-time algorithm for #CSP($\mathcal{F}$) when all three conditions are satisfied.

One open question is then to determine the decidability of our dichotomy criteria. Note that all the dichotomies discussed in the introduction are known to be decidable in NP, with many of them decidable in polynomial time. From the definitions of our dichotomy criteria, each of them requires one to check a condition on an infinitary object. Given a language $\mathcal{F}$ as the input, can we decide whether $\mathcal{F}$ satisfies all three conditions in finite time? If so, can we further show that the decision problem is in NP?